\documentclass[twoside,twocolumn,american,english,aps,pra]{revtex4-1}
\usepackage[T1]{fontenc}
\usepackage[latin9]{inputenc}
\usepackage{amsmath}
\usepackage{amssymb}
\usepackage{graphicx}
\usepackage{esint}

\usepackage{babel}

\begin{document}

\title{Surface code fidelity at finite temperatures}

\author{E. Novais}

\affiliation{Centro de Ci\^encias Naturais e Humanas, Universidade
  Federal do ABC, Santo Andr\a'{e}, SP, Brazil}

\author{A. J. Stanforth}

\affiliation{Department of Physics, University of Central Florida,
  Orlando, Florida 32816, USA}

\author{Eduardo R. Mucciolo}

\affiliation{Department of Physics, University of Central Florida,
  Orlando, Florida 32816, USA}

\begin{abstract}
We study the dependence of the fidelity of the surface code in the
presence of a single finite-temperature massless bosonic environment
after a quantum error correction cycle. The three standard types of
environment are considered: super-Ohmic, Ohmic, and sub-Ohmic. Our
results show that, for regimes relevant to current experiments,
quantum error correction works well even in the presence of
environment-induced, long-range interqubit interactions. A threshold
always exists at finite temperatures, although its temperature
dependence is very sensitive to the type of environment. For the
super-Ohmic case, the critical coupling constant separating high- from
low-fidelity decreases with increasing temperature. For both Ohmic and
super-Ohmic cases, the dependence of the critical coupling on
temperature is weak. In all cases, the critical coupling is determined
by microscopic parameters of the environment. For the sub-Ohmic case,
it also depends strongly on the duration of the QEC cycle.
\end{abstract}

\date{\today}

\maketitle

%%%%%%%%%%%%%%%%%%%%%%%%%%%%%%%%%%%%%%%%%%%%%%%%%%%%%%%%%%%%%%%%%%%%%%%%%%%%%%%
\section{Introduction}

A fundamental challenge to quantum information processing is
protection against detrimental effects of the environment
\cite{Unruh1995}. A milestone in addressing this problem was the
development of quantum error correction (QEC)
\cite{Steane1996,CS96}. In fact, it is believed that any practical
quantum information processing device will unavoidably contain some
sort of QEC \cite{Zurek2003}.

The main idea behind active QEC is to encode the information in a
region of the system's Hilbert space known as the logical subspace.
This region is chosen to be less vulnerable to the action of the
environment. However, during quantum evolution information can leak
out of this subspace. This leakage can be diagnosed by measuring some
observables in a process known as syndrome extraction. If an error is
detected by the syndromes, then a recovery operation is performed.
This sequence of actions, extracting the syndrome and a recovery
operation, can be called active QEC. It is clear that the QEC protocol
demands additional physical and computational resources, thus there is
a cost-benefit analysis that must be done. It is believed that there
is a particular noise strength below which the benefits of QEC
overcome its cost \cite{DAMB99}.

Although a large body of work has been devoted to including realistic
noise models in the analysis of the QEC efficacy
\cite{TerhalBurkard2005,AGP06,AKP2006,NB05,NMB07,NP09b,NERM10,YL15},
%there is little discussion on how the environmental temperature
%changes the noise threshold for {\it active} QEC. 
there is little discussion on the interplay between the environmental
temperature and the dynamics of the system affect the value of
threshold for {\it active} QEC. An important exception to this is
given by Brell and co-workers \cite{BBDFP2014} in the context of
topological quantum memories that are constantly monitored. In
contrast, for passive QEC it is well understood that a finite
environmental temperature radically affects the threshold
\cite{AH06,BT09,AFH09,PhysRevB.90.134302,arXiv:1411.6643}. In this
paper, we contribute to this theme by considering an environment whose
temperature changes due to the interaction with qubits when active QEC
is employed. We choose to evaluate the surface code performance
against the well-known pure dephasing model when a single logical
qubit is in an idle state against super-Ohmic and Ohmic bosonic
baths. In order to isolate the effects of finite temperature, we
consider a perfect syndrome extraction in a nonerror syndrome
evolution and assume all quantum gates and state preparations as
flawless. Hence, our thresholds should be regarded as an upper bound
to the real QEC threshold. Both analytical results and numerical
calculations are presented and we focus on temperatures and time
scales relevant to current experimental setups.

Our main result is that, in experimentally relevant regimes, an error
threshold always exists, but its dependence on temperature is not
universal. While for super-Ohmic environments the critical coupling
constant separating high fidelity and low fidelity behavior decreases
with increasing temperature, for Ohmic and sub-Ohmic environments the
dependence on temperature is weak. For the Ohmic cases, the critical
coupling depends primarily on microscopic parameters related to the
environment. For the sub-Ohmic case, it depends in addition on the
duration of the QEC cycle.

The paper is organized as follows. In Sec. \ref{sec:QEC} we provide a
brief description of the effect of quantum error correction in the
fidelity of a logical qubit coupled to an environment and the
evolution is not restricted to uncorrelated errors only. To account
for correlations, we consider environments composed of free boson. In
Sec. \ref{sec:finiteT} we develop a finite-temperature formulation for
the fidelity of the surface code in the presence of a bosonic
environment. The calculation of the fidelity after one QEC cycle is
mapped onto the calculation of certain expectation values of a
statistical spin model. The analysis that follows in
Sec. \ref{sec:puredephasing} is restricted to a realistic regime where
time-correlations (namely, thus developed inside the bosonic light
cone) predominate. Three significant cases are studied, namely,
super-Ohmic, Ohmic, and sub-Ohmic environments. Both analytical and
numerical results are presented. Concluding remarks and a summary are
presented in Sec. \ref{sec:conclusions}. Appendixes with detailed
results of calculations and a description of methods employed are
provided.

%%%%%%%%%%%%%%%%%%%%%%%%%%%%%%%%%%%%%%%%%%%%%%%%%%%%%%%%%%%%%%%%%%%%%%%%%%%%%%%
\section{Quantum error correction in the presence of correlated errors}
\label{sec:QEC}

%%%%%%%%%%%%%%%%%%%%%%%%%%%%%%%%%%%%%%%%%%%%%%%%%%%%%%%%%
%\subsection{Quantum error correction}

In order to focus on the effects of correlated errors and the ability
of QEC to tame the environmental degrees of freedom, we make two
simplifying assumptions. First, we assume that a quantum state can be
perfectly prepared. Hence, we consider that the environment and the
quantum system are disentangled at the beginning of the
evolution. While this assumption could in principle be relaxed, it
would result in a much more cumbersome calculation, that would obscure
the main effects that we want to discuss. Second, we assume that the
environment itself can be initially set to its lowest possible energy
state. This choice can easily be relaxed, but would lead to lower
threshold values. Even though we allow for the initial state of the
environment to be its ground state, we do not assume that it is
returned to the ground state at the end of the QEC cycle. Physically,
we are considering that the preparation of the initial state and the
environment could take a very long time (and we choose the best
possible preparation). When the system starts to evolve, the QEC
dynamics introduce a finite time scale that limits one's ability to
refrigerate the environment.

In order to clarify the notation and provide a self-contained
discussion, we start by giving a brief description of QEC. Following
the standard formulation of QEC, see Ref. \cite{JP-notes}, the unitary
evolution of a qubit and its environment in the interaction picture
can be described as
\begin{eqnarray}
\hat{U} \left|\psi\right\rangle \left|e_{0}\right\rangle & = & \hat{I}
\left|\psi\right\rangle \left|e_{I}\right\rangle +
\hat{X}\left|\psi\right\rangle \left|e_{X}\right\rangle +
\hat{Y}\left|\psi\right\rangle \left|e_{Y}\right\rangle \nonumber \\ &
& +\ \hat{Z}\left|\psi\right\rangle
\left|e_{Z}\right\rangle,
\label{eq:evolution2qec-1qubit-general}
\end{eqnarray}
where $\left|e_{\sigma}\right\rangle $ are environment states (in
general non orthogonal and non normalized) and $\hat{I}$, $\hat{X}$,
$\hat{Y}$, and $\hat{Z}$ are Pauli operators acting on the qubit. For
a system comprised of $n$ qubits, we can straightforwardly define an
expansion similar to Eq. (\ref{eq:evolution2qec-1qubit-general}),
namely,
\begin{equation}
\hat{U} \left|\psi\right\rangle \left|e_{0}\right\rangle = \sum_{a}
\hat{E}_{a}\left|\psi\right\rangle \left|e_{a}\right\rangle,
\label{eq:evolution-2qec-nqubits-general}
\end{equation}
where $\hat{E}_{a} \in \left\{ \hat{I},\hat{X},\hat{Y},\hat{Z}\right\}
^{\otimes n}$.

At the core of any QEC code is the choice of a particular subset
${\cal E} \subseteq \left\{ \hat{E}_{a}\right\} $, known as the
\emph{error set}, which the code can correct. The complementary set
${\cal \bar{E}}$ are the uncorrectable errors. It is therefore natural
to write the quantum evolution as
\begin{equation}
\hat{U} \left|\psi\right\rangle \left|e_{0}\right\rangle =
\sum_{a\in{\cal E}} \hat{E}_{a} \left|\psi\right\rangle
\left|e_{a}\right\rangle + \sum_{b\notin{\cal E}}
\hat{E}_{b}\left|\psi\right\rangle \left|e_{b}\right\rangle.
\end{equation}
The next step in a QEC protocol is the syndrome extraction, where a
set of observables corresponding to the projector ${\cal
  \hat{P}}_{\alpha}$ are measured in order to diagnose the errors and
then an appropriate recovery operation $\hat{R}_{\alpha}$ is chosen:
\begin{equation}
\hat{R}_{\alpha} \hat{{\cal P}}_{\alpha} \hat{U}
\left|\psi\right\rangle \left|e_{0}\right\rangle =
\left|\psi\right\rangle \left|e_{\alpha}\right\rangle +
\sum_{b\notin{\cal E}} \hat{R}_{\alpha} \hat{{\cal P}}_{\alpha}
\hat{E}_{b} \left|\psi\right\rangle \left|e_{b}\right\rangle .
\end{equation}
QEC is in essence a method to steer the quantum evolution of a qubit
system through a series of syndrome extractions. Even though QEC in
itself is not a perturbative method or description, its cost-benefit
analysis is usually done by performing a perturbative expansion in the
coupling between the environment and the system. To understand this
point, let us consider the fidelity of an initial state after a single
QEC step is performed and a syndrome $\alpha$ is detected. The
fidelity of the logical state in this case is given by
\begin{equation}
{\cal F}_{\alpha} = 1 - \frac{\left|\left|\sum_{b\notin{\cal E}}
  \hat{R}_{\alpha} \hat{{\cal P}}_{\alpha} \hat{E}_{b}
  \left|\psi\right\rangle \left|e_{b}\right\rangle
  \right|\right|}{\left|\left| \hat{R}_{\alpha} \hat{{\cal
      P}}_{\alpha} \hat{U} \left|\psi\right\rangle
  \left|e_{0}\right\rangle \right|\right|},
\end{equation}
where, for simplicity, we assumed that the environment states are
orthogonal to each other. After the syndrome is extracted only a
subset of terms in the Dyson series of the operator $\hat{U}$ is
kept. This expansion in the coupling with the environment is used in
many calculations of QEC. For instance, in the surface code it is an
essential ingredient in the understanding of minimal-weight matching
decodings. Hence, if
\begin{equation}
\left|\left| \sum_{b\notin{\cal E}} \hat{R}_{\alpha} \hat{{\cal
    P}}_{\alpha} \hat{E}_{b} \left|\psi\right\rangle
\left|e_{b}\right\rangle \right|\right| \ll \left| \left|
\hat{R}_{\alpha} \hat{{\cal P}}_{\alpha} \hat{U}
\left|\psi\right\rangle \left|e_{0}\right\rangle \right|\right|,
\end{equation}
a high fidelity can be achieved. Thus the choice of ${\cal E}$ and its
complement is a choice of the perturbative expansion imposed by the
error syndrome $\hat{{\cal P}}_{\alpha}$ extracted for a particular
evolution. Clearly, the fidelity can differ from unity due to
uncorrectable errors.

Choosing a recovery operation can be difficult in the surface code
\cite{PNTM2014}. There are strategies for choosing the most likely
$\hat{R}_{\alpha}$ for a certain syndrome. However, there is no
guarantee that the correct one is chosen. Hence the nonerror syndrome
turns out to be of special interest,
\begin{equation}
{\cal F}_{0} = 1 - \frac{\left|\left|\sum_{b\notin{\cal E}} \hat{{\cal
      P}}_{0} \hat{E}_{b} \left|\psi\right\rangle
  \left|e_{b}\right\rangle \right|\right|}{\left| \left| \hat{{\cal
      P}}_{0} \hat{U} \left|\psi\right\rangle \left|e_{0}\right\rangle
  \right|\right|}.
\end{equation}
It requires no recovery operation; thus ${\cal F}_{0}$ corresponds to
an intrinsic property of the error model. In this sense, it is
expected to provide an upper bound to the fidelity after a QEC cycle
with an arbitrary syndrome\cite{PNTM2014}. We restrict our analysis to
the nonerror syndrome case hereafter.

%%%%%%%%%%%%%%%%%%%%%%%%%%%%%%%%%%%%%%%%%%%%%%%%%%%%%%%%%%%%%
\subsection{Correlated and uncorrelated errors}

To discuss the concept of a threshold, we need to define a measure of
the noise strength. One way to do that is through the fidelity of a
\emph{single} physical qubit,
\begin{eqnarray}
{\cal F}_{\mbox{single qubit}} & = & \left\langle
e_{0}\right|\left\langle \psi\right|\hat{U}^{\dagger}
\left|\psi\right\rangle \left\langle \psi\right| \hat{U} \left|\psi
\right\rangle \left|e_{0}\right\rangle \nonumber \\ & = & \left\langle
e_{I}|e_{I}\right\rangle \nonumber \\ & = & 1 - p,
\label{eq:single-qubit-error-p}
\end{eqnarray}
where $p$ is named the \emph{single qubit error probability}. Thus, we
can rewrite Eq. (\ref{eq:single-qubit-error-p}) as
\begin{equation}
p = 1 - {\cal F_{\mbox{single qubit}}}.
\end{equation}
With this quantity, it is possible to define the concept of
\emph{uncorrelated errors}. For instance, for a two-qubit system, the
evolution with a nonerror syndrome is said to be uncorrelated if it is
possible to write the fidelity resulting from the QEC cycle as
\begin{eqnarray}
{\cal F}_{\mbox{two qubits}} & = & \left\langle
e_{0}\right|\left\langle \psi\right| \hat{U}^{\dagger}
\left|\psi\right\rangle \left\langle \psi\right| \hat{U}
\left|\psi\right\rangle \left|e_{0}\right\rangle ,\nonumber \\ & = &
\left(1-p\right)^{2}.
\end{eqnarray}
Conversely, when such decomposition of the noise evolution is not
possible, the problem is said to contain \emph{correlated errors}.

Most quantum error threshold discussions in the literature rely on the
existence of a \emph{single qubit error probability,} $p$, and,
explicitly or implicitly, rely on uncorrelated error
models. Furthermore, we note that the decompositions of
Eqs. (\ref{eq:evolution2qec-1qubit-general}) and
(\ref{eq:evolution-2qec-nqubits-general}) are, in general, only valid
for a single QEC step. The iteration of the process to the next QEC
step demands that the environment and the qubits be again
disentangled. Thus, any memory effects between QEC steps are formally
excluded in many discussions of the error threshold.

%%%%%%%%%%%%%%%%%%%%%%%%%%%%%%%%%%%%%%%%%%%%%%%%%%%%%%%%%%%%
\subsection{A microscopic model for correlated errors}

A paradigm model in the study of decoherence is the spin-boson model
for pure dephasing \cite{Weiss1999,Breuer2002}. The model consists of
free bosons coupled linearly to qubits and whose total
$\hat{H}=\hat{H}_{0}+\hat{H}_{{\rm int}}$ contains the free-boson term
\begin{equation}
\hat{H}_{0} = \sum_{{\bf k}} \omega_{{\bf k}}\, \hat{a}_{{\bf
    k}}^{\dagger} \hat{a}_{{\bf k}}
\label{eq:hamiltonian-0}
\end{equation}
and the qubit-boson interaction \cite[Chap. 4]{Breuer2002}
\begin{equation}
\hat{H}_{{\rm int}} = \lambda \sum_{{\bf r}} \hat{f} \left({\bf
  r}\right) \hat{\sigma}_{{\bf r}}^{x},
\label{eq:hamiltonian-1}
\end{equation}
where $\hat{\sigma}_{{\bf r}}^{x}$ is an $x$ spin operator for the
qubit located at site ${\bf r}$, with $\left[ \hat{a}_{{\bf
      k}},\hat{a}_{{\bf q}}^{\dagger} \right] =\delta_{{\bf k},{\bf
    q}}$, $\omega_{{\bf k}}$ defines the dispersion relation, and
\begin{equation}
\hat{f}\left({\bf r}\right) = \frac{(v/\omega_{0})^{D/2+s}}{L^{D/2}}
\sum_{{\bf k}\neq0}g_{{\bf k}}\, \left( e^{i{\bf k}\cdot{\bf r}}\,
\hat{a}_{{\bf k}}^{\dagger} + \mbox{H.c.} \right).
\label{eq:f-function}
\end{equation}
Here, $D$ is the number of spatial dimensions of the bath, $L$ is its
linear dimension, $\omega_{0}$ is a characteristic microscopic
frequency scale ($\hbar=1$), and $v$ is the bosonic velocity. In
Eq. (\ref{eq:hamiltonian-1}), $\lambda$ is the qubit-bath coupling
constant, which we separate from the form factor $g_{{\bf k}}$. For
convenience, the exponent $s$ is chosen such that $\hat{f}$ is
dimensionless and $\lambda$ has units of energy or frequency. 

It is straightforward to write the resulting evolution operator in
the interaction picture and in normal order,
\begin{equation}
\hat{U}\left(t\right) = \prod_{{\bf k}\neq0} e^{-\hat{{\cal G}}(t;{\bf
    k})}\, e^{-i\hat{\alpha}\,\left(t;{\bf k}\right)\, \hat{a}_{{\bf
      k}}^{\dagger}}\, e^{-i\hat{\alpha}^{*}\,\left(t;{\bf
    k}\right)\,\hat{a}_{{\bf k}}},
\end{equation}
where
\begin{widetext} 
\begin{equation}
\hat{\alpha} \left(t;{\bf k} \right) =
\frac{\lambda(v/\omega_{0})^{D/2+s}}{L^{D/2}} \sum_{{\bf r}}
\frac{g_{{\bf k}}}{\omega_{{\bf k}}} \sin \left( \omega_{k}t/2 \right)
\hat{\sigma}_{{\bf r},n}^{x}\, e^{i{\bf k}\cdot{\bf r}+i\omega_{{\bf
      k}}t/2}
\end{equation}
and 
\begin{equation}
\hat{{\cal G}}\left(t;{\bf k}\right) = \frac{\lambda^{2}}{4L^{D}}
(v/\omega_{0})^{D+2s} \int_{0}^{t} dt_{1} \int_{0}^{t} dt_{2}
\sum_{{\bf r},{\bf s}} \left| g_{{\bf k}} \right|^{2} e^{-i{\bf
    k}\cdot\left({\bf r}-{\bf s} \right)-i\omega_{{\bf
      k}}\left(t_{1}-t_{2}\right)}\, \hat{\sigma}_{{\bf r}}^{x}\,
\hat{\sigma}_{{\bf s}}^{x}\, \theta \left(t_{1}-t_{2} \right).
\end{equation}
\end{widetext}
Even though this model does not contain a full set of errors, it is
amenable to an exact and explicitly analytical description. Hence it
is well suited for exploring the effects of correlations, as well as
nonperturbative effects in QEC \cite{NB05}.

There are several possible regimes of correlations that can be
discussed using this model \cite{PNM2013}. They can be classified
according to the asymptotic behavior found after tracing out the
environment as follows.
\begin{enumerate}
\item Super-Ohmic: when some correlation functions of the system have
  an ultraviolet divergence in the cutoff frequency of the
  environment. Of course, there are no real divergences on a physical
  system, with the ultraviolet divergence just signaling that a more
  accurate description of the small-scale local physics of the qubit
  is missing from the model.
\item Ohmic: there are log divergences in the ultraviolet and in the
  infrared correlation functions. This is a more universal behavior,
  since results depend only on the logarithmic of the ultraviolet and
  infrared scales of the system.
\item Sub-Ohmic: all correlation functions of the system have a
  well-defined ultraviolet behavior, but some have infrared
  divergences.
\end{enumerate}

Both infrared and ultraviolet divergences can be troublesome to QEC,
but are amenable by suitable engineering of physical qubits. An
ultraviolet divergence signals that the qubit is strongly coupled to
the environment at high frequencies. This divergence is controlled by
form factors in the qubit design and therefore can be dealt with by
appropriate qubit engineering. Conversely, infrared divergences are
connected to long-distance correlations. This problem can be addressed
by better encoding designs. If we demand only local operations and
local communications between physical qubits, an infrared divergence
sets a limit on the number of qubits that one can have on the same
physical setup.

%%%%%%%%%%%%%%%%%%%%%%%%%%%%%%%%%%%%%%%%%%%%%%%%%%%%%%%%%%%%%%%%%%%%%%%%%%%%%%%%
\section{QEC with a bath at finite temperature}
\label{sec:finiteT}

We assume that the qubits can be prepared in the logical state $\left|
\psi \right\rangle$. In order to simplify the calculation and in
accordance with Eq. (\ref{eq:evolution2qec-1qubit-general}), we
consider the initial state of the bosonic environment to be the
vacuum, $\left| e_{0} \right\rangle = \left| 0 \right\rangle$. A mixed
initial state for the environment could also be considered, but would
make the notation and the calculation more complex, obscuring the
analysis. Thus, the qubits and the environment are initially in the
product state $\left| \psi, 0 \right\rangle$. In Appendix
\ref{initial_state} we discuss a possible initialization prescription.

After evolving for a time $\Delta$, the density matrix of the combined
system becomes
\begin{equation}
\hat{\rho} (\Delta) = \hat{U}(\Delta)\, |\psi,0\rangle\langle\psi,0|\,
\hat{U}^{\dagger}(\Delta).
\end{equation}
The next step is the syndrome extraction. We assume that the result of
this extraction is a nonerror. The occurrence of other types of
syndromes would introduce another layer of choices on the decoding
procedure, and therefore would potentially further reduce the
threshold (see discussion in Ref. \cite{PNTM2014}). Hence we
postselect the result of the syndrome in order to write the quantum
operation \cite{Breuer2002}
\begin{equation}
\Phi_{0} \left( \hat{\rho}_{s} \right) = \sum_{m} \hat{\cal P}_{0}\,
\hat{\rho}_{s}\, \hat{\cal P}_{0}^{\dagger},
\end{equation}
where $\hat{\rho}_{s}$ is a density matrix in the Hilbert space of the
qubits and the Kraus operator is the projector
\begin{equation}
\hat{\cal P}_{0} = \left| \bar{\uparrow} \right\rangle \left\langle
\bar{\uparrow} \right| + \left| \bar{\downarrow} \right\rangle
\left\langle \bar{\downarrow} \right|.
\end{equation}

A measurement can be understood as the selection of a pointer basis
due to the interaction of the measuring apparatus with another, fast
acting, environment \cite{Breuer2002}. Hence, during the time that the
syndrome is extracted, it is unphysical to regard the total system
(bosonic environment and the qubits) as isolated. The extraction of
the syndrome bound us to also discuss how the environment would behave
during this part of the evolution.

We cannot control the bosonic environment degrees of freedom, but it
is possible to place it into contact with an even larger reservoir.
This interaction can lead to a dissipative dynamics for the
environment that can help in reducing correlations and memory effects.
The qubits dynamics cannot affect the environment in any substantial
form during the interaction time. Hence, the quantum operation
that describes this evolution is
\begin{equation}
\Phi_{\beta} \left( \hat{\rho}_{e} \right) = \sum_{n} \hat{K}_{n}\,
\hat{\rho}_{e}\, \hat{K}_{n}^{\dagger},
\end{equation}
where $\hat{\rho}_{e}$ is a reduced density matrix in the environment
Hilbert space, the Kraus operators are
\begin{equation}
\hat{K}_{n} = \frac{e^{-\beta E_{n}/2}} {\sqrt{Z(\beta)}} \left | n
\rangle \langle n \right|,
\end{equation}
$E_{n}$ and $|n\rangle$ are the eigenvalues and eigenvectors of
$\hat{H}_0$, and, finally, the partition function
$Z(\beta)=\sum_{n}e^{-\beta E_{n}}$. Since $
\sum_n\hat{K}^{\dagger}_n\, \hat{K}_n \leq I$, the result of the quantum
operation has to be normalized and the density matrix after the
operation is given by
\begin{equation}
\tilde{\rho}_e(\beta) = \frac{\Phi_{\beta} \left( \hat{\rho}_{e}
  \right)} {\mathrm{tr}_e \left[ \Phi_{\beta} \left( \hat{\rho}_{e}
    \right) \right]}.
\end{equation}

Although it is tempting to associate $\beta$ with the inverse
temperature of the larger reservoir, it is straightforward to see that
this is an incorrect interpretation. To fully understand the physics
of $\hat{K}_n$, let us consider a few examples. The first case is the
action of $\Phi_\beta$ on the ``infinity temperature'' density matrix,
$\hat{\rho}_e(\infty) \equiv \sum_n \left | n \rangle \langle n
\right|$,
\begin{eqnarray}
\tilde{\rho}_e(\beta) & = & \frac{\Phi_{\beta} \left( \hat{\rho}_e
  (\infty) \right)} {\mathrm{tr}_e \left[ \Phi_{\beta} \left(
    \hat{\rho}_e (\infty)\right) \right]}, \nonumber\\ & = & \sum_n
\frac{e^{-\beta E_{n}}}{Z(\beta)} \left| n \rangle \langle n \right|.
\label{rho_beta}
\end{eqnarray}
The bosonic environment is brought from an ``infinity'' to a $1/\beta$
temperature. Now, if we apply $\Phi_{\beta}$ to Eq. (\ref{rho_beta}),
we obtain
\begin{equation}
\frac{\Phi_{\beta} \left( \hat{\rho}_e(\beta) \right)}{\mathrm{tr}_e
  \left[ \Phi_{\beta} \left( \hat{\rho}_e (\beta) \right) \right]} =
\sum_n \frac{e^{-2\beta E_{n}}}{Z(2\beta)} \left| n \rangle \langle n
\right|,
\label{rho_beta_2}
\end{equation}
thus corresponding now to an ensemble characterized by a even smaller
temperature, $1/(2\beta)$. In general, the operation $\Phi_e$ enhances
the probabilities of low-energy states in an statistical ensemble
instead of equilibrating it at a certain temperature. Therefore, it is
possible to call it a refrigeration or cooling mechanism. In Appendix
\ref{sec:cooling} we discuss a microscopic model that implements this
quantum operation.

Combining both quantum operations, $\Phi = \Phi_\beta \otimes
\Phi_{0}$, produces an action on the Hilbert space of the qubits and
the environment. This quantum operation is not normalized, since it is
not trace preserving. Thus the correct quantum evolution is given by
\begin{eqnarray}
\tilde{\rho}(\Delta) & = & \Phi \left( \hat{\rho}(\Delta) \right) \\ &
= & \frac{ \sum_{n} \hat{\cal P}_{0}\, \hat{K}_{n}\,
  \hat{\rho}(\Delta)\, \hat{K}_{n}^{\dagger}\, \hat{\cal P}_{0}}
{\text{tr} \left[ \sum_{n} \hat{\cal P}_{0}\, \hat{K}_{n}\,
    \hat{\rho}(\Delta)\, \hat{K}_{n}^{\dagger}\, \hat{\cal P}_{0}
    \right]}
\end{eqnarray}
and the reduced density matrix of the qubits is equal to
\begin{equation}
\tilde{\rho}_{s} \left(\Delta\right) = \text{tr}_{e} \left[
  \tilde{\rho} (\Delta) \right].
\end{equation}

For the qubits, the fidelity between the reduced density matrix at the
end of the QEC cycle and the initial density matrix is given by the
expression
\begin{equation}
{\cal F} = \text{tr}_{s} \left[ \tilde{\rho}_{s} \left(\Delta\right)
  \tilde{\rho}_{s} \left(0\right) \right],
\end{equation}
which can be rewritten as
\begin{equation}
{\cal F} = \frac{ \left\langle 0, \bar{\uparrow} \right|
  \hat{U}^{\dagger} \left(\Delta\right) \hat{\cal P}_{0} \left|
  \bar{\uparrow} \right\rangle \left\langle \bar{\uparrow} \right|
  e^{-\beta \hat{H}_{0}}\, \hat{\cal P}_{0}\,
  \hat{U}\left(\Delta\right) \left| 0, \bar{\uparrow} \right\rangle}
{\left\langle 0, \bar{\uparrow} \right| \hat{U}^{\dagger}
  \left(\Delta\right) \hat{\cal P}_{0}\, e^{-\beta \hat{H}_{0}}\,
  \hat{U}\left(\Delta\right) \left|0, \bar{\uparrow} \right\rangle},
\label{eq:FT}
\end{equation}
where we used the relation $\sum_{n} \hat{K}_{n}^{\dagger}\,
\hat{K}_{n} = \sum_n \frac{e^{-\beta E_{n}}}{Z(\beta)} \left| n
\rangle \langle n \right| = e^{-\beta \hat{H}_{0}}/Z(\beta)$.

The best possible scenario for QEC is when, at the end of the cycle,
the environment remains at zero temperature. This situation was
considered in Refs. \cite{NM2013,PNM2013,PNTM2014}. It corresponds to
forcefully setting the environment back to its ground state, hence
suppressing some correlations and memory effects. Even in this extreme
optimistic case, strong correlations among the qubits can still
persist, leading to a nontrivial threshold.

We can further simplify Eq.~(\ref{eq:FT}) by considering that
\begin{eqnarray}
{\rm tr} \left[ |\psi \rangle \langle \psi | \, \hat{\rho} (\Delta)
  \right] & = & \frac{e^{-\beta E_{0}}}{Z(\beta)}\, \langle \psi,0 |\,
\hat{U}^{\dagger} (\Delta;0)\, \hat{{\cal P}}_{0}\, |\psi \rangle
\nonumber \\ & & \times \langle \psi|\, \hat{{\cal P}}_{0}\,
\hat{U}(\Delta;\beta)\, |\psi,0 \rangle
\end{eqnarray}
and 
\begin{equation}
{\rm tr} \left[ \hat{\rho}(\Delta) \right] = \frac{e^{-\beta
    E_{0}}}{Z(\beta)}\, \langle \psi,0| \hat{U}^{\dagger} (\Delta;0)\,
\hat{{\cal P}}_{0}\, \hat{U} (\Delta;\beta)\, |\psi,0\rangle,
\end{equation}
where $\hat{U}(\Delta;\beta) = e^{-\beta\hat{H}_{0}}\,
\hat{U}(\Delta)\, e^{\beta\hat{H}_{0}}$ and we used $\hat{\cal{P}}_0^2
= \hat{\cal{P}}_0$. The end result is that the fidelity can be
rewritten as
\begin{equation}
{\cal F}_{0} = \frac{\langle\psi,0|\, \hat{U}^{\dagger}(\Delta;0)\,
  \hat{{\cal P}}_{0}\, |\psi\rangle\langle\psi|\, \hat{{\cal P}}_{0}\,
  \hat{U}(\Delta;\beta)\, |\psi,0\rangle} {\langle\psi,0
  |\hat{U}^{\dagger} (\Delta;0)\, \hat{{\cal P}}_{0}\,
  \hat{U}(\Delta;\beta)\, |\psi,0\rangle}.
\end{equation}
When we specialize to the pure dephasing bosonic model,
Eq. (\ref{eq:hamiltonian-1}), we obtain a compact expression for the
evolution operator
\begin{equation}
\hat{U}(\Delta;\beta) = \prod_{{\bf k}\neq0} e^{-\hat{{\cal
      G}}(\Delta;{\bf k})}\, e^{-i\hat{\alpha}(\Delta;{\bf
    k};\beta)\,\hat{a}_{{\bf k}}^{\dagger}}\,
e^{-i\hat{\alpha}^{\ast}(\Delta;{\bf k};\beta)\,\hat{a}_{{\bf k}}},
\end{equation}
where 
\begin{eqnarray}
\hat{\alpha} (\Delta;{\bf k};\beta) & = & \frac{\lambda}{L^{D/2}}
\left( \frac{v}{\omega_{0}} \right)^{D/2+s} \sum_{{\bf r}}
\frac{g_{{\bf k}}}{\omega_{{\bf k}}} \sin \left(
\frac{\omega_{k}\Delta}{2} \right) \nonumber \\ & \times &
\hat{\sigma}_{{\bf r},n}^{x}\, e^{i{\bf k}\cdot{\bf r}+i\omega_{{\bf
      k}}\left(\frac{\Delta}{2}+i\beta\right)}
\end{eqnarray}
and
\begin{eqnarray}
\hat{\alpha}^{*} (\Delta;{\bf k};\beta) & = & \frac{\lambda}{L^{D/2}}
\left( \frac{v}{\omega_{0}} \right)^{D/2+s} \sum_{{\bf r}}
\frac{g_{{\bf k}}^{*}}{\omega_{{\bf k}}} \sin \left(
\frac{\omega_{k}\Delta}{2} \right) \nonumber \\ & \times &
\hat{\sigma}_{{\bf r},n}^{x}\, e^{-i{\bf k}\cdot{\bf r}-i\omega_{{\bf
      k}}\left(\frac{\Delta}{2}+i\beta\right)}.
\end{eqnarray}
%

%%%%%%%%%%%%%%%%%%%%%%%%%%%%%%%%%%%%%%%%%%%%%%%%%%%%%%%%%%%5
\subsection{Surface code and the pure dephasing model}

The surface code \cite{DKLP2002} is regarded as a benchmark among the
QEC protocols \cite{FSG2009, Stephens2014}. It is implemented on a
two-dimensional array of qubits, greatly simplifying the design of
control and measurement circuits \cite{Jones2012}. In addition, all
the required interactions between the qubits are spatially
local. Finally, it has been estimated that it has a large noise
threshold in the absence of correlated errors \cite{WFH2011}.

\begin{figure}[ht]
\includegraphics[width=8cm]{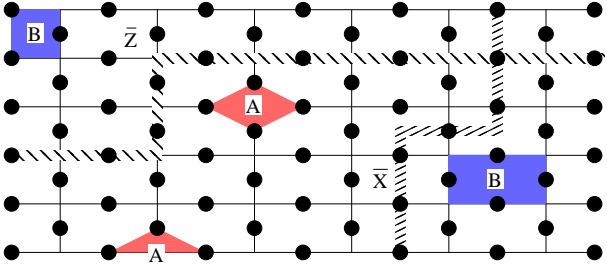}
\caption{(Color online) Surface code lattice and
  operators. Physical qubits are located at the edges (black
  circles). Some star $\hat{A}$ and plaquette $\hat{B}$ operators are
  indicated, as well, as some realizations of the logical operators
  $\hat{\bar{X}}$ and $\hat{\bar{Z}}$.}
\label{fig:lattice} 
\end{figure}

The surface code has the qubits located on the links of a
two-dimensional square lattice, as shown in
Fig. \ref{fig:lattice}. The quantum code is defined by two sets of
operators. The first set corresponds to local operators that define
the syndromes that have to be extracted at each QEC step. These are
four-body observables,
\begin{equation}
\hat{A}_{\lozenge} = \prod_{{\rm r}\in\lozenge} \hat{\sigma}_{{\rm
    r}}^{x}
\end{equation}
and 
\begin{equation}
\hat{B}_{\square} = \prod_{{\rm r}\in\square} \hat{\sigma}_{{\rm
    r}}^{z},
\end{equation}
where $\lozenge$ is a label for the positions of four qubits linked to
a vertex of the lattice (``star'') and $\square$ labels the positions
of four qubits in a plaquette. To diagnose the evolution of a single
logical qubit, all plaquette and star operators have to be measured in
order for the syndrome to be extracted.

%%%%%%%%%%%%%%%%%%%%%%%%%%%%%%%%%%%%%%%
%%%%%%%%%%%%%%%%%%%%%%%%%%%%%%%%%%%%%%

The second set corresponds to two extended operators that act on the
logical Hilbert space,
\begin{equation}
\hat{\bar{X}} = \prod_{{\rm r}\in\Gamma} \hat{\sigma}_{{\rm r}}^{x}
\end{equation}
and 
\begin{equation}
\hat{\bar{Z}} = \prod_{{\rm r}\in\Gamma^{\prime}} \hat{\sigma}_{{\rm
    r}}^{z},
\end{equation}
where $\Gamma$ is any path running from top to bottom of the lattice
(dual path) and $\Gamma^{\prime}$ is any path running from left to
right of the lattice (primal path), see Fig. \ref{fig:lattice}. Hence,
the logical Hilbert space is defined by the basis vectors
\begin{equation}
\left| \bar{\uparrow} \right\rangle = \frac{1} {\sqrt{N_{\lozenge}}}\,
\hat{G} \left| F_{z} \right\rangle
\end{equation}
and 
\begin{equation}
\left| \bar{\downarrow} \right\rangle = \hat{\bar{X}} \left|
\bar{\uparrow} \right\rangle,
\end{equation}
where $\hat{G} = \prod_{\lozenge} \left( 1 + \hat{A}_{\lozenge}
\right)$, $N_{\lozenge}=2^{n}$ is a normalization constant, $n$ is the
number of stars, and $\hat{\sigma}_{{\rm r}}^{z} \left| F_{z}
\right\rangle = \left| F_{z} \right\rangle$ for a qubit locate at
position ${\rm r}$. The nonerror syndrome projector in the logical
Hilbert space can then be written as
\begin{equation}
\hat{{\cal P}}_{0} = \left| \bar{\uparrow} \right\rangle \left\langle
\bar{\uparrow} \right| + \hat{\bar{X}} \left| \bar{\uparrow}
\right\rangle \left\langle \bar{\uparrow} \right| \hat{\bar{X}}.
\label{eq:P0}
\end{equation}

In order to investigate the error threshold for a pure dephasing
model, it is sufficient to consider that the system is initially
prepared in the logical state $\left|\bar{\uparrow}\right\rangle
$. Then, the fidelity for a nonerror evolution can be expressed as
\begin{eqnarray}
{\cal F}_{0} & = & \frac{{\cal A}} {{\cal A}+{\cal B}},
\end{eqnarray}
where we introduced the amplitudes
\begin{equation}
{\cal A} = \left\langle 0 \right|\left\langle \bar{\uparrow} \right|
\hat{U}^{\dagger} (\Delta,0) \left| \bar{\uparrow} \right\rangle
\left\langle \bar{\uparrow} \right |\hat{U}(\Delta;\beta) \left|
\bar{\uparrow} \right\rangle \left| 0 \right\rangle,
\label{eq:cal_A}
\end{equation}
and 
\begin{equation}
{\cal B} = \left\langle 0 \right| \left\langle \bar{\uparrow} \right|
\hat{U}^{\dagger}(\Delta,0)\, \hat{\bar{X}} \left| \bar{\uparrow}
\right\rangle \left\langle \bar{\uparrow} \right| \hat{\bar{X}}\,
\hat{U}(\Delta;\beta) \left| \bar{\uparrow} \right\rangle \left|0
\right\rangle.
\label{eq:cal_B}
\end{equation}
The evolution operators, the star operators, and the logical operators
in Eqs. (\ref{eq:cal_A}) and (\ref{eq:cal_B}) are diagonal in the $x$
basis of the qubits. Thus it is natural to rewrite the ferromagnetic
state as
\begin{equation}
\left| F_{z} \right\rangle = \frac{1}{2^{N/2}} \sum_{\sigma} \left|
\sigma \right\rangle,
\label{eq:Fz}
\end{equation}
where $\sigma$ labels the $2^{N}$ eigenstates of the operator
$\prod_{{\bf r}} \hat{\sigma}_{{\bf r}}^{x}$, namely, $\prod_{{\bf r}}
\hat{\sigma}_{{\rm r}}^{x} \left| \sigma \right\rangle = \pm \left|
\sigma \right\rangle$. Using Eq. (\ref{eq:Fz}) and assuming
$\omega_{{\bf k}}$ is isotropic in ${\bf k}$ \cite{footnote}, we can
write the amplitudes as
\begin{equation}
{\cal A} = \sum_{\sigma,\tau} e^{-\lambda^{2}{\cal H}(\Delta;\beta)}
\langle \tau| \hat{G} |\tau \rangle\langle \sigma| \hat{G}| \sigma
\rangle
\label{eq:calA}
\end{equation}
and 
\begin{equation}
{\cal B} = \sum_{\sigma,\tau} e^{-\lambda^{2}{\cal H}\Delta;\beta)}
\langle \tau | \hat{\bar{X}} \hat{G}| \tau \rangle \langle \sigma |
\hat{\bar{X}} \hat{G} | \sigma \rangle,
\label{eq:calB}
\end{equation}
where 
\begin{widetext} 
\begin{eqnarray}
{\cal H}(\Delta;\beta) & = & \frac{N}{2} F(\Delta;0;0) - \frac{1}{2}\,
F(\Delta;0;\beta) \sum_{{\bf r}} \sigma_{{\bf r}} \tau_{{\bf r}} +
\frac{1}{4} \sum_{{\bf r}\neq{\bf s}} \left[ F(\Delta;{\bf r} - {\bf
    s};0) \left( \tau_{{\bf r}} \tau_{{\bf s}} + \sigma_{{\bf r}}
  \sigma_{{\bf s}} \right) \right. \nonumber \\ & &
  \left. -\ F(\Delta;{\bf r}-{\bf s};\beta) \left( \sigma_{{\bf r}}
  \tau_{{\bf s}} + \tau_{{\bf r}} \sigma_{{\bf s}} \right) + i
  \Phi(\Delta;{\bf r}-{\bf s}) \left( \tau_{{\bf s}} - \sigma_{{\bf
      s}} \right) \left( \tau_{{\bf r}} + \sigma_{{\bf r}} \right)
  \right],
\label{eq:effective_hamiltonian}
\end{eqnarray}
\begin{equation}
F(\Delta;{\bf r};\beta) = \frac{(v/\omega_{0})^{D+2s}}{L^{D}}
\sum_{{\bf k}\neq0} \left| g_{{\bf k}} \right|^{2}
e^{-\beta\omega_{{\bf k}}} \left[ \frac{1-\cos(\omega_{{\bf k}}
    \Delta)} {\omega_{{\bf k}}^{\text{2}}} \right] \cos({\bf
  k}\cdot{\bf r}),
\label{eq:F(r)}
\end{equation}
and
\begin{equation}
\Phi(\Delta;{\bf r}) = \frac{(v/\omega_{0})^{D+2s}}{L^{D}}\sum_{{\bf
    k}\neq0} \left| g_{{\bf k}} \right|^{2} \left[ \frac{\omega_{{\bf
        k}} \Delta - \sin({\bf k}\Delta)}{\omega_{{\bf
        k}}^{2}} \right] \cos({\bf k}\cdot{\bf r}).
\label{eq:phi1(r)}
\end{equation}
Notice that both functions $F$ and $\Phi$ contain a dependence on
interqubit distance, but only the former depends on the environment
temperature.

\end{widetext}

Equations (\ref{eq:calA}) to (\ref{eq:phi1(r)}) are quite general.
They represent the mapping of the evaluation of the fidelity onto the
computation of expectation values of a classical spin system
comprising two coupled square-lattice layers and a complex Hamiltonian
${\cal H}$, with $\lambda^{2}$ playing the role of an effective
inverse temperature. Notice that the presence of the operator
$\hat{G}$ in the matrix elements entering in the expressions for
${\cal A}$ and ${\cal B}$, Eqs. (\ref{eq:calA}) and (\ref{eq:calB}),
constrains the sums over the variables $\sigma$ and $\tau$ to
configurations with positive plaquettes, namely, to configurations
with $\hat{A}_{\diamond}| \sigma \rangle = |\sigma\rangle$ and
$\hat{A}_{\diamond}| \tau \rangle = |\tau\rangle$. For configurations
containing negative plaquettes, the matrix elements are identically
zero.

At this point, in order to carry out a calculation of the fidelity, it
is necessary to consider a concrete example. We choose to discuss the
well-known pure dephasing decoherence model of
Refs.~\cite{Unruh1995,Breuer2002}. Thus we specialize our analysis to
situations comprising the following conditions:
\begin{enumerate}
\item a two-dimensional ($D=2$) environment; 
\item a linear dispersion relation, $\omega_{{\bf k}}=v|{\bf k}|$;
\item a coupling between the qubits and the environmental modes with a
  power-law behavior, $g_{{\bf k}}=\left|{\bf k}\right|^{s}$;
\item a bosonic ultraviolet cutoff $\Lambda$ for the qubit form factor
  $g_{{\bf k}}$ smaller than the environment's natural cutoff.
\end{enumerate}
With these conditions fulfilled and taking the infinite-volume limit,
we can rewrite the auxiliary functions in Eqs. (\ref{eq:F(r)}) and
(\ref{eq:phi1(r)}) as
\begin{eqnarray}
F(\Delta;{\bf r};\beta) & = & \frac{1}{\pi} \frac{1}{\omega_{0}^{2}\,
  (\omega_{0} \Delta)^{2s}} \int_{0}^{\infty}dx\, x^{2s-1}\, J_{0}
\left(\frac{\left|{\bf r}\right|x}{v\Delta} \right) \nonumber \\ & &
\times\ (1-\cos x)\,
e^{-\left(\beta+\frac{1}{v\Lambda}\right)x/\Delta},
\end{eqnarray}
and 
\begin{eqnarray}
\Phi(\Delta;{\bf r}) & = & \frac{1}{\pi} \frac{1}{\omega_{0}^{2}
  (\omega_{0} \Delta)^{2s}} \int_{0}^{\infty}dx\, x^{2s-1}\, J_{0}
\left( \frac{\left|{\bf r}\right|x}{v\Delta} \right)\nonumber \\ & &
\times\ (x-\sin x)\, e^{-x/(v\Delta\Lambda)},
\end{eqnarray}
where $J_{0}(z)$ is the zeroth order Bessel function. The parameter
$s$ defines the correlation regime of the model: $s>0$ corresponds to
a super-Ohmic , $s=0$ to an Ohmic, and $s<0$ to a sub-Ohmic environment.

Following the well-known phenomenology of the single-qubit case,
$\beta$ defines the thermal correlation time
\cite{Breuer2002}. Whenever $\Delta\ll\beta$, the system is in the
vacuum regime and systematic corrections can be evaluated in powers of
$\Delta/\beta$. The opposite case is the thermal regime,
$\beta<\Delta$, which has not been previously studied in the context
of the surface code. It is important to note that finite temperature
means simply that even though the environment is prepared at zero
temperature, the external cooling mechanism cannot suppress the
bosonic excitations during the evolution of the system. The functions
$F(\Delta;{\bf r};\beta)$ and $\Phi(\Delta;{\bf r})$ in different
regimes are presented in Appendix \ref{sec:Coupling-constants}.

Finally, in order to numerically evaluate the threshold we need to
make some additional choices. Guided by the most recent experimental
developments of superconducting qubits \cite{Chen2014}, which are good
candidates for implementing the surface code, we assume a certain range
of values for the model's microscopic parameters.

\begin{enumerate}
\setcounter{enumi}{4}

\item It is reasonable to assume that in a running QEC protocol the
  environmental temperature for superconducting qubits is of the order
  of a few millikelvin. Hence, we set $\beta =
  \frac{\hbar}{k_BT}\approx 10^{-9}$ s.

\item It is also reasonable to consider that the QEC period $\Delta$
  is of the order of $100$ ns. Therefore, we only consider the thermal
  regime $\beta\ll\Delta$.

\item Furthermore, the distance between nearest neighbors qubits, $a$,
  in an superconducting qubit array is likely to be of the order of
  $10^{-5}$ m.

\end{enumerate}

All the above choices are very reasonable. We note that the thermal
regime is likely to be applicable to physical implementations other
than the superconducting qubits as well.

The only parameter that is difficult to estimate is the velocity of
the bosonic environment. Its value can vary by several orders of
magnitude depending on the dominant physical environment. For
instance, a typical phonon velocity in solid-state substrates is
$v=10^{3}$ m/s; however, electromagnetic fluctuations propagate with
$v=10^{8}$ m/s. Roughly, for every power on $10$ in the bosonic
velocity, the number of qubits in the {\it time-like} cone increases
by $10$. Hence, all qubits in an experimental set up would be
time-like correlated for the latter case since $v \Delta \gg \left|
{\bf r} \right|$. This is likely less so for the phononic environment,
but time-like correlations should predominate. Thus, in the following
we assume $v \Delta > \left| {\bf r} \right|$.

%%%%%%%%%%%%%%%%%%%%%%%%%%%%%%%%%%%%%%%%%%%%%%%%%%%%%%%%%%%%%%%%%%%%%%%%%%%%%%%%
\section{QEC threshold for the pure dephasing model}
\label{sec:puredephasing}

%%%%%%%%%%%%%%%%%%%%%%%%%%%%%%%%%%%%%%%%%%%%%%%%%%%%%%%%%%%%%%%%%%%%%%%%%
\subsection{Super-Ohmic environment with $s=1/2$}
\label{sec:super-Ohmic}

\begin{figure}[ht]
\includegraphics[width=0.9\columnwidth]{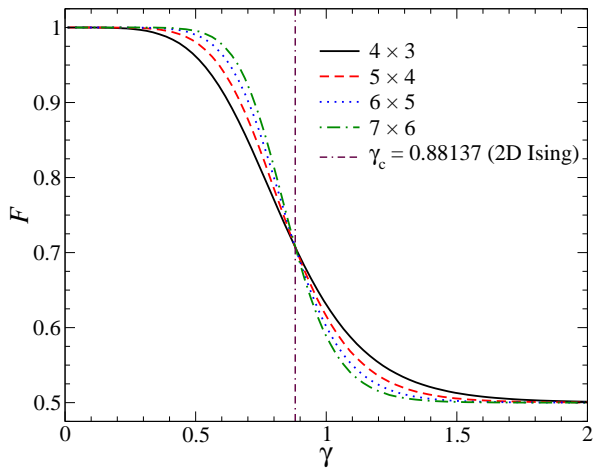}
\caption{(Color online) Fidelity in the presence of a super-Ohmic
  environment ($s=1/2$) in the thermal regime for several system
  sizes. Data obtained by averaging over $10^9$ Monte Carlo
  steps. Here, $\gamma = \lambda^{2}F(\Delta;0;\beta)$. The dashed
  line marks the phase transition point of an Ising model on a square
  lattice with nearest-neighbor interactions.}
\label{fig:Fidelity-super-Ohmic}
\end{figure}

The $s=1/2$ environment can describe an acoustic phonon bath or an
electromagnetic environment \cite{Mahan2000}. Using the expressions
for the coupling functions defined in Eqs. (\ref{eq:F(r)}) and
(\ref{eq:phi1(r)}) presented in Table \ref{table} of Appendix
\ref{sec:Coupling-constants}, we clearly see that for a super-Ohmic
bath $F\left(\Delta;0;0\right)$ diverges with the ultraviolet cutoff
and $F\left(\Delta;0;\beta\right)$ diverges with the inverse of the
temperature in the thermal regime. Moreover, the ratio of any of other
coupling function present in Eq. (\ref{eq:effective_hamiltonian}) by
one these two diverging couplings tends to zero. Hence, since
$F(\Delta,{\bf r},0)$ and $F(\Delta,{\bf r},\beta)$ can be made of the
same order in the thermal regime, we can simplify the statistical
model and keep only the leading interaction, namely,
\begin{equation}
{\cal H}_{\rm super} (\Delta;\beta) = -\, \frac{1}{2}\,
F(\Delta;0;\beta) \sum_{{\bf r}} \sigma_{{\bf r}}\, \tau_{{\bf r}}\:,
\label{eq:Hsuper-Ohmic}
\end{equation}
in order to describe the effect of a super-Ohmic environment on the
fidelity. The purely local (yet constrained) spin model defined by the
effective two-body interaction of Eq. (\ref{eq:Hsuper-Ohmic}) can be
solved exactly by introducing an auxiliary plaquette variable
\begin{equation}
\mu_{{\bf r}-\delta}\, \mu_{{\bf r}+\delta} = \sigma_{{\bf r}}\,
\tau_{{\bf r}},
\end{equation}
with ${\bf r}\pm\delta$ labeling the plaquettes that share the link
where the qubit ${\bf r}$ is located. The statistical sum over the
$\mu$ variables in Eqs. (\ref{eq:calA}) and (\ref{eq:calB}) is
unconstrained \cite{PNTM2014}. Thus the introduction of plaquette spin
variables maps the problem onto a standard two-dimensional Ising model
with boundary fields. In the thermodynamic limit,
$N\rightarrow\infty$, there is a well-known critical coupling
\cite{Salinas}
\begin{equation}
\lambda_{c} = \left[ \frac{\ln \left( 1 + \sqrt{2} \right)} {F \left(
    \Delta;0;\beta \right)} \right]^{1/2},
\label{eq:critical-coupling-super-Ohmic}
\end{equation}
separating a region where the fidelity is equal to 1
$(\lambda<\lambda_{c})$ from a region where the fidelity goes to $1/2$
$(\lambda>\lambda_{c})$. We confirm this analytical result performing
a standard Monte Carlo simulation, as shown in
Fig. \ref{fig:Fidelity-super-Ohmic}. The random walk is performed in
the mass field variables, while the energy updates are computed using
the original spin variables. The numerical simulation clearly
indicates the QEC threshold predicted by
Eq. (\ref{eq:critical-coupling-super-Ohmic}). We stress that this
result is fundamentally different from previous results by the authors
\cite{NM2013,PNTM2014,PNM2013}, where the limit $\beta\to\infty$ was
taken and therefore the nearest-neighbor coupling in the statistical
spin model was the dominant term.

A remarkable feature of Eq. (\ref{eq:critical-coupling-super-Ohmic}) is
that the critical coupling for the threshold in a super-Ohmic
environment has a square root dependence with the inverse of the
temperature but is independent of the QEC time $\Delta$,
\begin{equation}
\lambda_{c} \propto \omega_{0}\, \sqrt{\omega_{0}\, \beta}.
\end{equation}
Therefore, for any value of the microscopic coupling with the
environment, there will always be a sufficiently low temperature for
which the fidelity of the qubit will be 1.

%%%%%%%%%%%%%%%%%%%%%%%%%%%%%%%%%%%%%%%%%%%%%%
%%%%%%%%%%%%%%%%%%%%%%%%%%%%%%%%%%%%%%%%%%%%%%%

The introduction of the nearest-neighbor coupling to ${\cal H}_{\rm
  super}$ with the real coupling function $F$, as prescribed by
Eq. (\ref{eq:effective_hamiltonian}), does not dramatically change
these results. However, the addition of the term with the imaginary
part $\Phi$ could, in principle, introduce enough oscillations to
remove the threshold. Thus we explore numerically the stability of the
threshold against the introduction of an imaginary part between
nearest neighbors by studying the modified Hamiltonian
\begin{eqnarray}
{\cal H}_{\rm super}(\Delta;\beta) & = & -\, \frac{1}{2}\,
F(\Delta;0;\beta) \sum_{{\bf r}} \sigma_{{\bf r}}\, \tau_{{\bf r}}
\nonumber \\ & & +\ \frac{i}{4}\, \Phi(\Delta;d) \sum_{\langle {\bf
    r},{\bf s}\rangle } \left( \tau_{{\bf s}}-\sigma_{{\bf s}}\right)
\left(\tau_{{\bf r}} + \sigma_{{\bf r}} \right), \nonumber \\
\label{eq:Hsuper_imag}
\end{eqnarray}
where $d=a/\sqrt{2}$. For time-like correlations and in the thermal
regime we expect $\eta \equiv \Phi(\Delta;d)/F(\Delta;0;\beta) \sim
\beta/\Delta \ll 1$.

\begin{figure}[ht]
\includegraphics[width=.95\columnwidth]{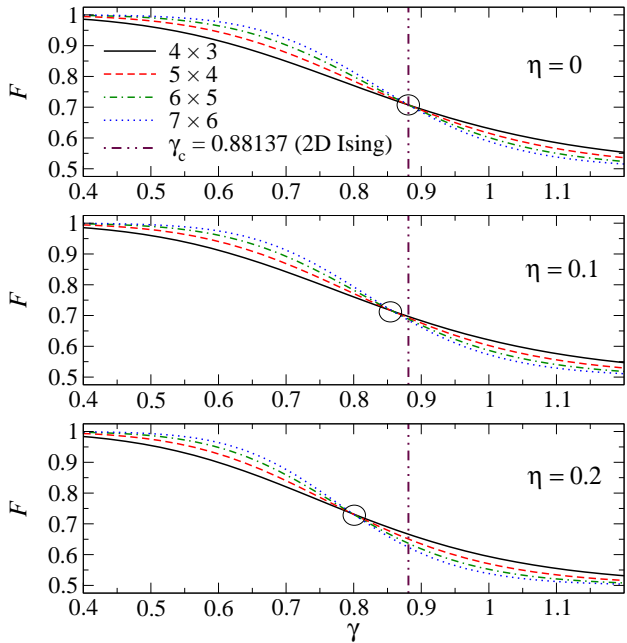}
\caption{(Color online) Fidelity of the surface code in the presence
  of a super-Ohmic environment ($s=1/2$) for several system sizes when
  a small imaginary nearest-neighbor interaction is present; see
  Eq. (\ref{eq:Hsuper_imag}). Data points obtained using Binder's
  method. Here, $\gamma = \lambda^{2}F(\Delta;0;\beta)$ and $\eta =
  \Phi(\Delta;d) / F(\Delta;0;\beta)$. The dashed line marks the phase
  transition point of an Ising model in a square lattice with
  nearest-neighbor interactions. The circle marks the threshold
  position.}
\label{fig:Fidelity-super-Ohmic-binder}
\end{figure}

The complex interaction in Eq. (\ref{eq:Hsuper_imag}) prevents the use
of the Monte Carlo method. We employ instead Binder's recursive method
to compute the amplitudes ${\cal A}$ and ${\cal B}$ \cite{KB72,GB90},
as explained in Appendix \ref{sec:Binder_method}. The results for the
fidelity are presented in Fig. \ref{fig:Fidelity-super-Ohmic-binder}
and show that a shift toward lower thresholds occurs. Hence we
conclude that the QEC threshold for the super-Ohmic environment, with
$s=1/2$, is mildly robust against small deviations from the asymptotic
model defined by Eq. (\ref{eq:Hsuper-Ohmic}). A threshold continues to
exists, but it is lowered due to the coherence oscillations caused by
the imaginary effective interaction term in the statistical model.

%%%%%%%%%%%%%%%%%%%%%%%%%%%%%%%%%%%%%%%%%%%%%%%
%%%%%%%%%%%%%%%%%%%%%%%%%%%%%%%%%%%%%%%%%%%%%%%

%%%%%%%%%%%%%%%%%%%%%%%%%%%%%%%%%%%%%%%%%%%%%%%%%%%%%%%%%%%%%%%%%%%%%%%%
\subsection{Ohmic environment}
\label{sec:Ohmic}

The Ohmic environment corresponds to $s=0$. Long-range correlations
are ubiquitous to this environment, hence it cannot be discussed in
the same manner as the super-Ohmic case (see Appendix
\ref{sec:Coupling-constants}).

Some analytical development can be made if we consider the limit where
all qubits interact with each other with the same strength (thus
taking the logarithmic interaction as a constant): $F \left(
\Delta,{\bf r}, \beta\right) = \bar{F}$ and $\Phi \left( \Delta,{\bf
  r} \right) = \bar{\Phi}$, with $\bar{F} \approx \bar{\Phi} \sim
1/\omega_0^2$. Physically, this corresponds to the distance between
the qubits being smaller than the thermal coherence length. In such a
simplified model, the Hamiltonian can be rewritten as
\begin{eqnarray}
{\cal H} \left(\Delta,\beta\right) & = & -\, \frac{1}{2}\, \Delta F
\sum_{{\bf r}} \sigma_{{\bf r}} \tau_{{\bf r}} + \frac{\bar{F}}{4}
\left[ \sum_{{\bf r}} \left( \sigma_{{\bf r}} - \tau_{{\bf r}}\right)
  \right]^{2} \nonumber \\ & & +\ i \frac{\bar{\Phi}}{4} \left[
  \sum_{{\bf r}} \left( \sigma_{{\bf r}} - \tau_{{\bf r}}\right)
  \right] \left[ \sum_{{\bf r}} \left( \sigma_{{\bf r}} + \tau_{{\bf
      r}}\right) \right] \\ & = & -\, \frac{1}{2}\, \Delta F
\sum_{{\bf r}} \sigma_{{\bf r}} \tau_{{\bf r}} + \frac{\bar{F}}{4}
(m_\sigma - m_\tau)^2 \nonumber \\ & & +\ i \frac{\bar{\Phi}}{4}
(m_\sigma - m_\tau) (m_\sigma + m_\tau),
\label{eq:longrangemodel}
\end{eqnarray}
where $\Delta F\equiv F(\Delta;0;\beta)-\bar{F}$, $m_\sigma \equiv
\sum_{\bf r} \sigma_{\bf r}$ and $m_\tau \equiv \sum_{\bf r} \tau_{\bf
  r}$. It is straightforward to show that, in the absence of QEC, this
model causes a reduction of the fidelity proportional to the square of
the number of qubits \cite{Unruh1995,Breuer2002}. The use of QEC
changes this scenario quite dramatically. The logical states of the
surface code have a finite probability amplitude in most total
magnetization sectors $(m_\sigma,m_\tau)$. Hence, for nearly every
qubit configuration without a logical error one can find another
configuration with a logical error and with the same value for the
difference $m_\sigma - m_\tau$. As a consequence, although the
effective Hamiltonian in Eq. (\ref{eq:longrangemodel}) contains
long-range interactions, they do not distinguish between
configurations with and without logical errors and the threshold is
controlled by the local term proportional to $\Delta F$.

\begin{figure}
\includegraphics[width=0.9\columnwidth]{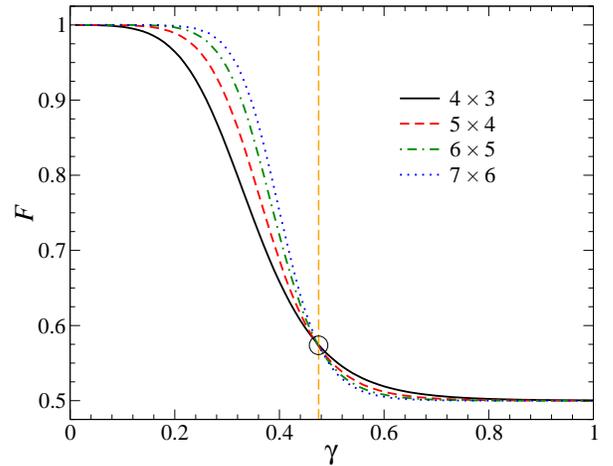}
\caption{(Color online) Fidelity of the surface code in the presence
  of an Ohmic bath, as given by the statistical model of
  Eq. (\ref{eq:longrangemodel}), for different lattice sizes. Here,
  $\gamma = \lambda^{2} \Delta F$, $\bar{\Phi}=0$ and $\bar{F} = 0.72
  \Delta F$. Data obtained by averaging over $10^9$ Monte Carlo
  steps. The location of the transition point, indicated by the
  crossing point surrounded by a circle, is marked by the dashed
  line.}
\label{fig:Ohmic-bath}
\end{figure}

We evaluate numerically the fidelity for the statistical model of
Eq. (\ref{eq:longrangemodel}) using the Monte Carlo method but
disregarding the imaginary part of the interaction. (Unfortunately
Binder's method is no longer practical when interactions go beyond
nearest neighbors.) We take $\bar{F} = 0.72 \Delta F$. The results are
presented in Fig. \ref{fig:Ohmic-bath} and indicate the existence of a
threshold for
\begin{equation}
\lambda_{c} \approx \left[ \frac{0.475}{\Delta
    F\left(\Delta;0;\beta\right)} \right]^{\frac{1}{2}},
\label{eq:critical_coupling_Ohmic}
\end{equation}
thus corresponding to a reduction of about $1/4$ of the super-Ohmic
value in Eq. (\ref{eq:critical-coupling-super-Ohmic}).

%%%%%%%%%%%%%%%%%%%%%%%%%%%%%%%%%%%%%%%%%%%%%
%%%%%%%%%%%%%%%%%%%%%%%%%%%%%%%%%%%%%%%%%%%%%

The most important difference to the super-Ohmic case is not the
numerical value in the coupling constant, but rather the
insensitivity to changes in temperature: the threshold depends on
temperature only through a logarithm,
\begin{equation}
\lambda_{c} \propto \omega_{0} \left| \ln \left( \frac{\Delta}{\beta}
\right) \right|^{-\frac{1}{2}}.
\end{equation}
As a result, in practice, only the microscopic frequency scale
$\omega_0$ determines whether a particular realization of the surface
code is above or below the threshold for a given QEC time $\Delta$.

%%%%%%%%%%%%%%%%%%%%%%%%%%%%%%%%%%%%%%%%%%%%%%%%%%%%%%%%%%%%%%%%%%%%%%%%%%%%%%%%
\subsection{Sub-Ohmic environment with $s=-1/2$}
\label{sec:sub-Ohmic}

A sub-Ohmic environment introduces correlations among very distant
qubits. In particular, for the case of $s=-1/2$, as presented in
Appendix \ref{sec:Coupling-constants}, we find that correlations are
weakly dependent on qubit-qubit distances and environment
temperature. Hence we can revive the analytical discussion used for
the Ohmic environment. Considering again the thermal regime and the
numerical estimates discussed in Sec. \ref{sec:Ohmic}, namely, $
\beta\ll\Delta$ and $v\beta\ll\left|{\bf r}\right|<v\Delta$, we find
that $\Delta F\approx \Delta/\omega_0$ and temperature independent up
to subleading logarithmic corrections. Thus, the critical coupling is
controlled by the microscopic characteristic frequency scale
$\omega_0$ and the QEC time, namely,
\begin{equation}
\lambda_c \propto \sqrt{\frac{\omega_0}{\Delta}}.
\end{equation}
%

%%%%%%%%%%%%%%%%%%%%%%%%%%%%%%%%%%%%%%%%%%%%%%%%%%%%%%%%%%%%%%%%%%%%%%%%%%%%%%%%
\section{Discussion and Summary}
\label{sec:conclusions}

It is unavoidable that during its quantum evolution a system will get
entangled with its environment. This entanglement can be understood as
an effective temperature that characterizes the system's reduced
density matrix. To make this point clear, let us consider the simple
example of a single qubit interacting with a bosonic environment
through the pure bit-flip model \cite{Breuer2002}. If the combined
system plus environment starts in the pure state $\left| \uparrow
\right \rangle_{z} \otimes \left| 0 \right \rangle $, and we use as
the effective Hamiltonian only the site-diagonal term in
Eq. (\ref{eq:effective_hamiltonian}), the fidelity of this single
qubit can be written as
\begin{equation}
{\cal F} \left(\Delta,\beta\right) = \frac{1} {1+\bar{M}_{x}},
\end{equation}
where $\bar{M}_{x} = \tanh \left[ \lambda^{2}
  \frac{F\left(\Delta;0;\beta\right)} {2} \right]$. The fidelity of
this qubit is a smooth function of $\lambda$, going continuously from
$1$ to $1/2$.

The $\bar{M}_{x}$ function can be understood as the mean magnetization
of a fictitious statistical mechanics problem of a qubit in the
presence of a magnetic field, $h = \frac{F\left(\Delta;0;\beta\right)}
{2}$, at a temperature $1/\lambda^{2}$. Notice that the actual degrees
of freedom of the statistical mechanics problem are not the original
qubit variables $\sigma$ and $\tau$, but instead the square of their
difference, namely,
\begin{equation}
\mu = 1 - \frac{1}{2} \left(\sigma-\tau\right)^{2}.
\end{equation}
In the spin-boson model literature \cite{Leggett:1987:1}, time
intervals where $\sigma=\tau$ are called ``sojourns''; while for
$\sigma=-\tau$ they are called ``blips''. Transitions between sojourns
and blips correspond to flips of the spin variable $\mu$.

This simple picture is precisely what we generalize in this paper. We
consider the fidelity of a single logical qubit coupled to a bosonic
environment through a pure bit-flip interaction. After tracing the
environment the dynamical problem can once again be mapped onto an
effective thermodynamics problem. The remarkable feature of QEC is to
transform a crossover into a true phase transition in the limit of a
logical qubit with an infinite number of physical qubits. The
microscopic parameters that define the transition point yield the
intrinsic noise threshold value for the code (which is independent of
decoding errors, see discussion in Ref. \cite{PNTM2014}). Ideally it
would be preferable to have simultaneously bit-flips and phase flips
in an error model; however, this is not fundamental to deduce the
existence or not of a threshold.

There are many regimes to consider, but it is very likely that
physical realizations of a quantum memory will be in the thermal,
$\beta\ll\Delta$, and time-correlated, $\left|{\bf r}\right|<v\Delta$,
regimes. For this range of parameters, our main conclusion is that the
surface code in a super-Ohmic environment always has a noise threshold
and the critical value of the qubit-environment coupling constant goes
as $\lambda_{c} \propto \omega_{0} \sqrt{\omega_{0}\beta}$, where
$\beta$ is the inverse temperature of the environment at the end of
the QEC cycle. Therefore, for the super-Ohmic case, it is always
possible to place the system below the noise threshold by reducing the
environmental temperature. In contrast, for Ohmic environments,
$\lambda_{c}$ is a weak function of temperature and only microscopic
parameters play a relevant role in determining whether QEC protects or
not the logical qubit state. For sub-Ohmic environments, $\lambda_c$
is also approximately temperature independent, but in addition to
depending on microscopic scales, it is inversely proportional to the
QEC cycle duration.

These results are overall reassuring and indicate that there is no
fundamental limitation to the existence of a noise threshold for the
surface code in the presence of bosonic environments after a single
QEC cycle. We are currently investigating the effects on the fidelity
of errors correlated over multiple cycles.

%In summary, we provide an exact mapping for the fidelity of a single
%logical qubit protected by the surface code in a finite temperature
%bosonic environment into an effective statistical mechanics
%problem. We specialize the calculation to physically relevant regions
%of the parameter space and evaluate the threshold as a function of the
%microscopic parameters of the model and the environment temperature
%after a QEC time $\Delta$.

%%%%%%%%%%%%%%%%%%%%%%%%%%%%%%%%%%%%%%%%%%%%%%%%%%%%%%%%%%%%%%%%%%%%%%%%%%%%%%%
\acknowledgments

This work was supported in part by the National Science Foundation
through Grant No. CCF-1117241, FAPESP Grant No. 2014/26356-9, and
INCT-IQ.

%%%%%%%%%%%%%%%%%%%%%%%%%%%%%%%%%%%%%%%%%%%%%%%%%%%%%%%%%%%%%%%%%%%%%%%%%%%%%%%
\appendix

\section{Initializing the state}
\label{initial_state}

Suppose that the environment is controlled by the Hamiltonian
$\hat{H}_{0}$. The Hamiltonian of the combined qubit-environment
system is
\begin{equation}
\hat{H} = \hat{H}_{0} + \sum_{{\bf r}} \left[ \hat{f} \left({\bf
    r}\right) \hat{\sigma}_{{\bf r}}^{z} - h\, \hat{\sigma}_{{\bf
      r}}^{z} \right],
\label{eq:Htotal}
\end{equation}
where $\hat{f}\left({\bf r}\right)$ represents the interaction between
the qubit at position ${\bf r}$ and the environment and $h$ is an
external field. The density matrix of the system at thermal
equilibrium reads
\begin{equation}
\hat{\rho} \left(\beta,h\right) = \frac{e^{-\beta \hat{H}}}{Z},
\end{equation}
where $Z$ is the partition function, $Z = \text{tr} \left[ e^{-\beta
    \hat{H}} \right]$.

The Hamiltonian in Eq. (\ref{eq:Htotal}) is diagonal in the qubit
space. Assume that $h$ is large and the qubits are frozen in the $+z$
direction. Then,
\begin{equation}
\hat{\rho} \left( \beta, h \to\infty \right) = \frac{e^{-\beta \left[
      \sum_{k} \omega_{k}\, \hat{a}_{k}^{\dagger}\, \hat{a}_{k} +
      \lambda \sum_{n} \hat{f} \left( n \right) \right]}} {Z} \otimes
\left| F \rangle \langle F \right|,
\end{equation}
where $\left|F\right>$ is the ferromagnetic $z$ state of the
qubits. Using Eq.~(\ref{eq:hamiltonian-0}), it is natural to define
$\hat{b}_{k} = \hat{a}_{k} + \alpha_{k}/\omega_{k}$ in order to
rewrite the bosonic Hamiltonian as
\begin{equation}
\omega_{k}\, \hat{b}_{k}^{\dagger}\, \hat{b}_{k} = \omega_{k}\,
\hat{a}_{k}^{\dagger}\, \hat{a}_{k} + \alpha_{k}\,
\hat{a}_{k}^{\dagger} + \alpha_{k}^{*}\, \hat{a}_{k} + \frac{ \left|
  \alpha_{k} \right|^{2}} {\omega_{k}}
\end{equation}
and the density matrix as
\begin{equation}
\hat{\rho} \left( \beta, h \to\infty \right) = \frac{ e^{-\beta \left[
      \sum_{k} \omega_{k}\, \hat{b}_{k}^{\dagger}\, \hat{b}_{k}
      \right]}} {\bar{Z}} \otimes \left( \left| F \rangle \langle F
\right| \right),
\end{equation}
where $\bar{Z}$ is the partition function for the new $\hat{b}_{k}$
bosons. Raking the temperature to zero ($\beta \to \infty$), we arrive
at
\begin{equation}
\hat{\rho} \left( \beta \to\infty, h \to\infty \right) = \left|
\bar{0} \rangle \langle \bar{0} \right| \otimes \left| F \rangle
\langle F \right|,
\end{equation}
where $|\bar{0}\rangle$ is the ground state of the $\hat{b}_{k}$
bosons. This ground state is a coherent state of the original
$\hat{a}_{k}$ bosons,

\begin{equation}
\hat{a}_{k} | \bar{0} \rangle = - \lambda \left| {\bf k} \right|^{s-1}
\frac{\sum_{n} e^{-i{\bf k} \cdot {\bf r}_{n}}} {vL^{D/2}}\, | \bar{0}
\rangle.
\end{equation}
However, if the qubits do not form a dense set with respect to the
bosonic environment, then, in the limit $L\to\infty$,
\begin{equation}
|\bar{0} \rangle = |0 \rangle,
\end{equation}
and we obtain the state
\begin{equation}
\hat{\rho} \left( \beta \to\infty, h \to\infty \right) = |0 \rangle
\langle 0| \otimes |F \rangle \langle F |.
\end{equation}
Finally, assuming the ability of instantaneous (faster than the
environment inverse cutoff) and flawless gates, the initial state can
be prepared as
\begin{eqnarray}
\hat{\rho}_{0} & = & |0\rangle \langle 0| \otimes G\, |F \rangle
\langle F| G \\ & = & |0 \rangle \langle 0| \otimes \left|
\bar{\uparrow} \right\rangle \left\langle \bar{\uparrow} \right|.
\end{eqnarray}
%

%%%%%%%%%%%%%%%%%%%%%%%%%%%%%%%%%%%%%%%%%%%%%%%%%%%%%%
\section{Microscopic Cooling Mechanism}
\label{sec:cooling}

A microscopic description of the cooling process of the free bosonic
environment coupled to the qubits proceeds as follows. The relation
of the bosonic environment and an external reservoir can be described
by the usual damped harmonic oscillator master equation
\cite{Breuer2002}. For an illustrative example of this microscopic
description, consider qubits inside an electromagnetic cavity. The
modes inside the cavity constitute the correlated
environment. However, there are electromagnetic modes outside the
cavity as well. These external modes can damp the modes inside the
cavity.

For a given bosonic mode, the master equation, after the usual
Born-Markov approximations, is given by
\begin{eqnarray}
\frac{d}{dt} \hat{\rho}_{{\bf k}} & = & -i \omega_{{\bf k}} \left[
  \hat{a}_{{\bf k}}^{\dagger}\, \hat{a}_{{\bf k}}, \hat{\rho}_{{\bf
      k}} \right] \nonumber \\ & & +\ \gamma_{\bf k} \left( N_{\bf k}
+ 1 \right) \left( \hat{a}_{{\bf k}}\, \hat{\rho}_{{\bf k}}\,
\hat{a}_{{\bf k}}^{\dagger} - \frac{1}{2} \hat{a}_{{\bf
    k}}^{\dagger}\, \hat{a}_{{\bf k}}\, \hat{\rho}_{{\bf k}} -
\frac{1}{2} \hat{\rho}_{{\bf k}}\, \hat{a}_{{\bf k}}^{\dagger}\,
\hat{a}_{{\bf k}} \right) \nonumber \\ & & +\ \gamma_{\bf k} N_{\bf k}
\left( \hat{a}_{{\bf k}}^{\dagger}\, \hat{\rho}_{{\bf k}}\,
\hat{a}_{{\bf k}} - \frac{1}{2} \hat{a}_{{\bf k}}\, \hat{a}_{{\bf
    k}}^{\dagger}\, \hat{\rho}_{{\bf k}} - \frac{1}{2}
\hat{\rho}_{{\bf k}}\, \hat{a}_{{\bf k}}\, \hat{a}_{{\bf
    k}}^{\dagger}\right),
\label{master-1}
\end{eqnarray}
where $N_{\bf k} = \left[\exp \left(\tilde{\beta} \omega_{{\bf
      k}}\right) -1 \right]^{-1}$, $\gamma_{\bf k}$ is the damping
rate, and we defined the inverse temperature $\tilde{\beta} = 1 /
\tilde{T}$ ($k_{B}=1$). In order to maximize the cooling and be
compatible with the initial state chosen for the bosonic environment
(see Appendix \ref{initial_state}), the external reservoir should be
at its lowest possible temperature, $\tilde{T} = 0$.

If we evoke the usual assumption that decoherence is much faster than
dissipation, we can focus on solving the master equation for the
populations, known as Pauli master equation. For $\tilde{T}=0$ it is
simply
\begin{equation}
\frac{d}{dt} P_{\bf k}(n,t) = \gamma_{\bf k} \left[ (n+1) P_{\bf k}
  (n+1,t) -n P_{\bf k}(n,t) \right],
\label{master-2}
\end{equation}
where $P_{\bf k}(n,t)= \langle {\bf k}; n| \hat{\rho}_{\omega_{{\bf
      k}}}(t) |{\bf k};n\rangle$ and $n$ denotes the number of ${\bf
  k}$ modes \cite{Breuer2002}.

These coupled differential equations are easily solvable if we
consider that the original populations are only sparsely nonzero,
i.e., if $P_{\bf k}(n,0) \neq 0$, then $P_{\bf k}(n\pm1,0) = 0$.
Considering that syndrome extraction takes a time $t = \epsilon$, the
initial population of mode ${\bf k}$ is reduced to
\begin{equation}
P_{\bf k}(n, \epsilon) = e^{-\gamma_{\bf k}\, n\, \epsilon}\, P_{\bf
  k}(n,0).
\end{equation}
It is also reasonable to assume that the damping rate is a function of
the energy of the bosonic mode. A simple choice is to make it a linear
relation,
\begin{equation}
\gamma_{\bf k}\, n\, \epsilon = \langle n| \beta\, \omega_{\bf k}\,
\hat{a}^\dagger_{\bf k}\, \hat{a}_{\bf k} |n \rangle.
\label{def_beta}
\end{equation}
This corresponds physically to having higher frequencies coupled more
strongly to the external reservoir than lower ones. Therefore, a given
environmental mode with initial density matrix
\begin{equation}
\hat{\rho}_{\bf k} (0) = \sum_{n,m} w_{n,m} |{\bf k}; n \rangle
\langle {\bf k};m|,
\end{equation}
evolves towards
\begin{equation}
\hat{\rho}_{{\bf k}} (\epsilon) = \sum_{n} w_{n,n}\, e^{-\beta\,
  \omega_{\bf k}\, \hat{a}^\dagger_{\bf k}\, \hat{a}_{\bf k}} |{\bf
  k}; n \rangle \langle {\bf k};n|.
\end{equation}

These considerations hold for all environmental modes ${\bf
  k}$. Finally, if we assume that decoherence would quickly destroy
the coherences between different environmental modes, we find that the
initial density matrix of the environment,
\begin{equation}
\hat{\rho}_{e} (0) = \sum_{i,j} w_{i,j}\, | i \rangle \langle j|,
\end{equation}
evolves towards the density matrix
\begin{equation}
\hat{\rho}_{e} (\epsilon) = \sum_{i} w_{i,i}\, e^{-\beta \hat{H}_0} |i
\rangle \langle i|.
\end{equation}
That corresponds to the quantum operation $\Phi_{\beta} \left(
\hat{\rho}_{e} (0)\right)$, where $\beta$ is defined as a function of
the damping rates of the environment, Eq. (\ref{def_beta}). The
particular case of $\beta=0$ corresponds to having no damping, hence
describing a situation where the environment has a unitary evolution
during the syndrome extraction.

%%%%%%%%%%%%%%%%%%%%%%%%%%%%%%%%%%%%%%%%%%%%%%%%%%%%%
\section{Coupling constants for some environments}
\label{sec:Coupling-constants}

The evaluation of Eqs. (\ref{eq:F(r)}) and (\ref{eq:phi1(r)}) is a
straightforward but long task. Closed forms can only be found for
special values of $s$. Here we present results for some representative
cases and for different environments. As discussed in the main text,
the inverse temperature $\beta$ defines the thermal coherence time,
creating two limiting regimes. For the quantum vacuum regime we can
assume $\Delta/\beta\ll1$ to evaluate the integrals. Conversely, for
the thermal regime we can assume that $\beta/\Delta\ll1$. Finally,
during the evaluation of Eqs. (\ref{eq:F(r)}) and (\ref{eq:phi1(r)})
it is assumed that all distances $\left|{\bf r}\right|$ are much
larger than $\mbox{max}\{v\beta,\Lambda^{-1}\}$. The results are
presented in Table \ref{table}.

%%%%%%%%%%%%%%%%%%%%%%%%%%%%%%%%%%%%%%%%%%%%%%%%%%%%%%%%%%%%
\begin{widetext}

\begin{table}[h]

\begin{tabular}{|c|c|c|c|}
\hline 
 & {\small{}}%
\begin{tabular}{c}
{\small{}super-Ohmic}\tabularnewline
{\small{}($s=\frac{1}{2})$}\tabularnewline
\end{tabular} & {\small{}}%
\begin{tabular}{c}
{\small{}Ohmic}\tabularnewline
{\small{}($s=0)$}\tabularnewline
\end{tabular} & {\small{}}%
\begin{tabular}{c}
{\small{}sub-Ohmic}\tabularnewline
{\small{}($s=-\frac{1}{2})$}\tabularnewline
\end{tabular}\tabularnewline
\hline 
\hline 
{\small{}$F(\Delta;0;0)$} & {\small{}}%
\begin{tabular}{c}
\tabularnewline
{\small{}$\frac{v\Lambda}{\pi\omega_{0}^{3}}$}\tabularnewline
\tabularnewline
\end{tabular} & {\small{}}%
\begin{tabular}{c}
\tabularnewline
\selectlanguage{american}%
{\small{}$\frac{1}{\pi}\frac{1}{\omega_{0}^{2}}\ln\left(v\Lambda\Delta\right)$}\selectlanguage{english}%
\tabularnewline
\tabularnewline
\end{tabular} & {\small{}}%
\begin{tabular}{c}
\tabularnewline
{\small{}$\frac{\Delta}{2\omega_{0}}$}\tabularnewline
\tabularnewline
\end{tabular}\tabularnewline
\hline 
{\small{}$\Phi(\Delta;{\bf r}\neq0)$} & {\small{}}%
\begin{tabular}{c}
\tabularnewline
{\small{}$\frac{v}{\pi\omega_{0}^{3}}\frac{\theta(v\Delta-\left|{\bf r}\right|)}{\sqrt{v^{2}\Delta^{2}-\left|{\bf r}\right|^{2}}}$}\tabularnewline
\tabularnewline
\end{tabular} & {\small{}}%
\begin{tabular}{c}
\tabularnewline
{\small{}$\frac{1}{\pi\omega_{0}^{2}}\left[\frac{\pi}{2}\theta(v\Delta-\left|{\bf r}\right|)+\right.$}\tabularnewline
{\small{}$\left.\arcsin\left(\frac{v\Delta}{\left|{\bf r}\right|}\right)\theta(\left|{\bf r}\right|-v\Delta)\right]$}\tabularnewline
\tabularnewline
\end{tabular} & {\small{}}%
\begin{tabular}{c}
{\small{}$\frac{\Delta}{\pi\omega_{0}}\ln\left[\sqrt{\left(\frac{v\Delta}{{\bf r}}\right)^{2}-1}+\frac{v\Delta}{{\bf r}}\right.$}\tabularnewline
{\small{}$\left.-\sqrt{1-\left(\frac{{\bf r}}{v\Delta}\right)^{2}}\right]\theta\left(v\Delta-{\bf r}\right)$}\tabularnewline
\tabularnewline
\end{tabular}\tabularnewline
\hline 
{\small{}$F_{\mbox{thermal}}(\Delta;0;\beta)$} & {\small{}}%
\begin{tabular}{c}
\tabularnewline
{\small{}$\frac{1}{\pi\omega_{0}^{3}\beta}$}\tabularnewline
\tabularnewline
\end{tabular} & {\small{}}%
\begin{tabular}{c}
\tabularnewline
{\small{}$\frac{1}{\pi}\frac{1}{\omega_{0}^{2}}\ln\left(\frac{\Delta}{\beta}\right)$}\tabularnewline
\tabularnewline
\end{tabular} & {\small{}}%
\begin{tabular}{c}
\tabularnewline
{\small{}$\frac{\Delta}{\pi\omega_{0}}\left(\frac{\pi}{2}+\frac{\beta}{\Delta}\ln\frac{\beta}{\Delta}\right)$}\tabularnewline
\tabularnewline
\end{tabular}\tabularnewline
\hline 
{\small{}$F_{\mbox{thermal}}(\Delta;{\bf r}\neq0;\beta)$} & {\small{}}%
\begin{tabular}{c}
\tabularnewline
\tabularnewline
{\small{}$\frac{v}{\pi\omega_{0}^{3}}\left[\frac{1}{\left|{\bf r}\right|}-\frac{\theta(\left|{\bf r}\right|-v\Delta)}{\sqrt{\left|{\bf r}\right|^{2}-v^{2}\Delta^{2}}}\right]$}\tabularnewline
\tabularnewline
\tabularnewline
\end{tabular} & {\small{}}%
\begin{tabular}{c}
\tabularnewline
\tabularnewline
{\small{}$\frac{1}{\pi\omega_{0}^{2}}\left[\mbox{arccosh}\left(\frac{v\Delta}{\left|{\bf r}\right|}\right)\theta(v\Delta-\left|{\bf r}\right|)\right.$}\tabularnewline
{\small{}$\left.+\frac{v\beta}{\sqrt{\left|{\bf r}\right|^{2}-\left(v\Delta\right)^{2}}}\theta\left(\left|{\bf r}\right|-v\Delta\right)-\frac{v\beta}{\left|{\bf r}\right|}\right]$}\tabularnewline
\tabularnewline
\tabularnewline
\end{tabular} & {\small{}}%
\begin{tabular}{c}
\tabularnewline
{\small{}$\frac{\Delta}{\pi\omega_{0}}\left\{ \left[\left(\frac{\pi}{2}-\frac{\left|{\bf r}\right|}{v\Delta}\right)+\right.\right.$}\tabularnewline
{\small{}$\left.\mbox{\ensuremath{\frac{\beta}{\Delta}}arccosh}\left(\frac{v\Delta}{\left|{\bf r}\right|}\right)\right]\theta(v\Delta-\left|{\bf r}\right|)$}\tabularnewline
{\small{}+$\left[\arcsin\frac{v\Delta}{\left|{\bf r}\right|}+\right.$}\tabularnewline
{\small{}$\left.\left.\sqrt{\left(\frac{\left|{\bf r}\right|}{v\Delta}\right)^{2}-1}-\frac{\left|{\bf r}\right|}{v\Delta}\right]\theta(\left|{\bf r}\right|-v\Delta)\right\} $}\tabularnewline
\tabularnewline
\end{tabular}\tabularnewline
\hline 
\end{tabular}

\caption{\label{table} Coupling constants for different noise regimes. }
\end{table}

\end{widetext}
%%%%%%%%%%%%%%%%%%%%%%%%%%%%%%%%%%%%%%%%%%%%%%%%%%%%%%%%%%%%%%%

%%%%%%%%%%%%%%%%%%%%%%%%%%%%%%%%%%%%%%%%%%%%%%%%%%%%%%%%%%%%%%%%%%%%%%%%%%%%%%%
\section{Mass field formulation}

\begin{figure}[h]
\includegraphics[width=8cm]{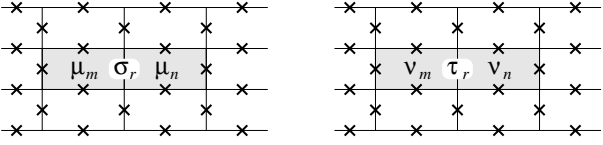} 
\caption{Mass field variable composition.}
\label{fig:mass-field-mapping}
\end{figure}

One of the main difficulties in evaluating Eqs. (\ref{eq:calA}) and
(\ref{eq:calB}) numerically is to enforce the positive star
constraints. An efficient method to enforce the constraint is to
introduce auxiliary plaquette variables, the so-called mass fields
\cite{PNTM2014}, for the bulk qubits in the surface code,
\begin{equation}
\sigma_{\mathbf{r}} = \mu_{\mathbf{m}} \mu_{\mathbf{n}},
\end{equation}
and 
\begin{equation}
\tau_{\mathbf{r}} = \nu_{\mathbf{m}} \nu_{\mathbf{n}},
\end{equation}
where $\mathbf{m}$ and $\mathbf{n}$ denote the plaquettes that share
the edge where the spin site $\mathbf{r}$ is located (see Fig.
\ref{fig:mass-field-mapping}). For qubits on the top and bottom
boundaries, we follow the discussion in Ref. \cite{NM2013} and use
instead
\begin{equation}
\sigma_{\mathbf{r}} = \mu_{\mathbf{m}}\alpha_{t}
\end{equation}
or 
\begin{equation}
\sigma_{\mathbf{r}} = \mu_{\mathbf{m}} \alpha_{b}
\end{equation}
and 
\begin{equation}
\tau_{\mathbf{r}} = \nu_{\mathbf{m}} \beta_{t}
\end{equation}
or 
\begin{equation}
\tau_{\mathbf{r}} = \nu_{\mathbf{m}} \beta_{b},
\end{equation}
with $\alpha_{t},\alpha_{b},\beta_{t},\beta_{b}=\pm1$.

%%%%%%%%%%%%%%%%%%%%%%%%%%%%%%%%%%%%%%%%%%%%%%%%%
%%%%%%%%%%%%%%%%%%%%%%%%%%%%%%%%%%%%%%%%%%%%%%%%%

The introduction of mass fields automatically enforce positive stars:
$\langle\mu|\hat{G}|\mu\rangle = \langle\nu|\hat{G}|\nu\rangle = 1$
for all $\mu$ and $\nu$. In addition,
\begin{eqnarray}
\langle\sigma | \hat{X}\hat{G}| \sigma\rangle & = & \alpha_{t}
\alpha_{b}
\end{eqnarray}
and 
\begin{eqnarray}
\langle\tau| \hat{X}\hat{G} | \tau\rangle & = & \beta_{t} \beta_{b}.
\end{eqnarray}

In the mass field variables it is evident that the effective energy of
the qubit configurations that contribute to the amplitudes in Eqs.
(\ref{eq:calA}) and (\ref{eq:calB}) obeys some symmetry
properties. For instance, time-reversal symmetry still holds,
\begin{equation}
E(\mu,\nu;\alpha,\beta) = E(-\mu,-\nu;-\alpha,-\beta),
\end{equation}
as well as complex conjugation through the exchange of mass-field and
boundary-field variables,
\begin{equation}
E(\mu,\nu;\alpha,\beta) = [E(\nu,\mu;\beta,\alpha)]^{\ast}.
\end{equation}

In addition to automatically enforcing the positive star constraints,
the mass fields are also very useful to deal with on-site and
nearest-neighbor interactions in the original qubits. If we restrict
ourselves to this particular case of nearest neighbor, and define
\begin{equation}
J = \frac{F\left(\Delta,a,\beta\right) + i \Phi \left(\Delta,a\right)}
{F\left(\Delta,0,\beta\right)},
\end{equation}
we find that the energy can be written as
\begin{widetext}
\begin{align*}
E(\mu,\nu;\alpha, \beta) & = E_{\mathrm{bulk}}(\mu,\nu) +
E_{\mathrm{top}}(\mu,\nu;\alpha_{t},\beta_{t}) +
E_{\mathrm{bottom}}(\mu,\nu;\alpha_{b},\beta_{b})
\end{align*}
where 
\begin{equation}
E_{\mathrm{bulk}}(\mu,\nu) = \frac{1}{2} \left[ N -
  \sum_{\langle\mathbf{m},\mathbf{n}\rangle} \mu_{\mathbf{m}}
  \mu_{\mathbf{n}} \nu_{\mathbf{m}} \nu_{\mathbf{n}} \right] +
J^{\ast} \sum_{\langle\langle\mathbf{m},\mathbf{n}\rangle\rangle}
\mu_{\mathbf{m}} \mu_{\mathbf{n}} + J
\sum_{\langle\langle\mathbf{m},\mathbf{n}\rangle\rangle}
\nu_{\mathbf{m}} \nu_{\mathbf{n}} - \mathrm{Re} \{J\} \left[
  \sum_{\langle\mathbf{m},\mathbf{n},\mathbf{m}'\rangle}
  \mu_{\mathbf{m}} \mu_{\mathbf{n}} \nu_{\mathbf{n}} \nu_{\mathbf{m}'}
  \right],
\end{equation}
\begin{eqnarray}
E_{\mathrm{top}} (\mu,\nu;\alpha_{t},\beta_{t}) & = & -\frac{1}{2}
\left[ \alpha_{t} \beta_{t} \sum_{\mathbf{m}\in y=N_{y}}
  \mu_{\mathbf{m}} \nu_{\mathbf{m}} \right] + J^{\ast} \alpha_{t}
\left[ \sum_{\mathbf{m}\in y=N_{y},2\leq x\leq N_{x}-1}
  \mu_{\mathbf{m}} + \frac{1}{2} (\mu_{1,N_{y}} + \mu_{N_{x},N_{y}})
  \right] \nonumber \\ & & +\ J \beta_{t} \left[ \sum_{\mathbf{m}\in
    y=N_{y},2\leq x\leq N_{x}-1} \nu_{\mathbf{m}} + \frac{1}{2}
  (\nu_{1,N_{y}} + \nu_{N_{x},N_{y}}) \right] \nonumber \\ & &
-\ \mathrm{Re} \{J\} \left[ \alpha_{t}
  \sum_{\langle\mathbf{m},\mathbf{n}\rangle\in y=N_{y}}
  \nu_{\mathbf{m}} \nu_{\mathbf{n}} \mu_{\mathbf{n}} + \beta_{t}
  \sum_{\langle\mathbf{m},\mathbf{n}\rangle\in y=N_{y}}
  \mu_{\mathbf{m}} \mu_{\mathbf{n}} \nu_{\mathbf{n}} \right],
\end{eqnarray}
and 
\begin{eqnarray}
E_{\mathrm{bottom}}(\mu,\nu;\alpha_{b},\beta_{b}) & = & -\frac{1}{2}
\left[ \alpha_{b} \beta_{b} \sum_{\mathbf{m}\in y=1} \mu_{\mathbf{m}}
  \nu_{\mathbf{m}} \right] + J^{\ast} \alpha_{b} \left[
  \sum_{\mathbf{m}\in y=1,2\leq x\leq N_{x}-1} \mu_{\mathbf{m}} +
  \frac{1}{2} (\mu_{1,1} + \mu_{N_{x},1}) \right] \nonumber \\ & &
+\ J \beta_{b} \left[ \sum_{\mathbf{m}\in y=1,2\leq x\leq N_{x}-1}
  \nu_{\mathbf{m}} + \frac{1}{2} (\nu_{1,1} + \nu_{N_{x},1}) \right]
\nonumber \\ & & -\ \mathrm{Re} \{J\} \left[ \alpha_{b}
  \sum_{\langle\mathbf{m},\mathbf{n}\rangle\in y=1} \nu_{\mathbf{m}}
  \nu_{\mathbf{n}} \mu_{\mathbf{n}} + \beta_{b}
  \sum_{\langle\mathbf{m},\mathbf{n}\rangle\in y=1} \mu_{\mathbf{m}}
  \mu_{\mathbf{n}} \nu_{\mathbf{n}}\right],
\end{eqnarray}
\end{widetext}
where $N_x$ and $N_y$ indicate the horizontal and vertical number of
plaquettes in the surface code lattice.

There are two features that complicate any numerical calculation: the
appearance of three-body interactions and the next-to-nearest neighbor
interactions. Both of these features make any recursive computation
very difficult and any Monte-Carlo simulation less efficient (if at
all possible, due to the presence of an imaginary coupling).

In terms of mass fields and boundary fields, the targets of the
calculation are the quantities
\begin{equation}
{\cal B} = \frac{1}{Z} \sum_{\alpha_{t},\alpha_{b}=\pm1}
\sum_{\beta_{t},\beta_{b}=\pm1} \sum_{\{\mu\}} \sum_{\{\nu\}} e^{-\xi
  E(\mu,\nu;\alpha,\beta)} \alpha_{t} \alpha_{b} \beta_{t}\beta_{b}
\end{equation}
and 
\begin{equation}
Z = \sum_{\alpha_{t},\alpha_{b}=\pm1} \sum_{\beta_{t},\beta_{b}=\pm1}
\sum_{\{\mu\}} \sum_{\{\nu\}} e^{-\xi E(\mu,\nu;\alpha,\beta)},
\end{equation}
where $\xi \equiv \lambda^{2} F\left(\Delta,0,\beta\right)$ is a
fictitious temperature. It is straightforward to prove that $Z$ and
${\cal B}$ are real quantities. Furthermore, from
$Z\left(\gamma\right)$ we can compute the expectation value of the
effective energy and corresponding heat capacity,
\begin{equation}
E = - \frac{d\ln Z}{d\xi},
\end{equation}
and 
\begin{equation}
C = -\xi^{2} \frac{dE}{d\xi}.
\end{equation}
Using the auxiliary function 
\begin{eqnarray}
c(\alpha,\beta) & \equiv & \sum_{\{\mu\}} \sum_{\{\nu\}} e^{-\xi
  E(\mu,\nu;\alpha,\beta)},
\end{eqnarray}
we can write 
\begin{eqnarray}
Z & = & 2 \left[ c(+,+,+,+) + 2 c(+,+,+,-) \right. \nonumber \\ & &
  +\ 2 c(+,-,+,+) + c(+,+,-,-) \nonumber \\ & & +\ \left. c(+,-,+,-) +
  c(+,-,-,+) \right]
\end{eqnarray}
and 
\begin{eqnarray}
{\cal B} & = & \frac{2}{Z} \left[ c(+,+,+,+) - 2 c(+,+,+,-)
  \right. \nonumber \\ & & -\ 2 c(+,-,+,+) + c(+,+,-,-) \nonumber \\ &
  & + \ \left. c(+,-,+,-) + c(+,-,-,+) \right],
\end{eqnarray}
where we used the time-reversal symmetry of $E(\mu,\nu;\alpha,\beta)$
to reduce the number of terms.

%%%%%%%%%%%%%%%%%%%%%%%%%%%%%%%%%%%%%%%%%%%%%%%%%%%%%%%%%%%%%%%%%%%%%%%%%%%%%%%
\section{Binder's recursive method for the surface
code with nearest-neighbors interactions}
\label{sec:Binder_method}

We can extend Binder's recursive method \cite{KB72,GB90} for computing
the partition function of the Ising model in a two-dimensional square
lattice with nearest-neighbor interactions to the effective
statistical model of Eq. (\ref{eq:effective_hamiltonian}). Two
modifications are necessary: (i) to consider two spins per site and
(ii) to introduce auxiliary variables in the recursive steps.

Suppose we start with a square lattice of dimensions $N_{x}\times
N_{y}$ ($N_{x}$ columns and $N_{y}$ rows). To each lattice site on a
row we associate two binary numbers $(s_{x},r_{x})$, with $s_{x}=0,1$,
$r_{x}=0,1$ and $x=1,\ldots,N_{x}$. We can then index the state of the
spins in a lattice row by two integers $(\mathbf{s},\mathbf{r})$,
where
\begin{equation}
\mathbf{s} = s_{N} \times 2^{N-1} + s_{N-1} \times 2^{N-2} + \ldots +
s_{2} \times 2^{1} + s_{1} \times 2^{0}
\end{equation}
and 
\begin{equation}
\mathbf{r} = r_{N} \times 2^{N-1}+r_{N-1} \times 2^{N-2} + \ldots +
r_{2} \times 2^{1}+r_{1} \times2^{0}.
\end{equation}

The numbers $\mathbf{s}$ and $\mathbf{r}$ are related to the mass
field variables $\mu$ and $\nu$ by
\begin{eqnarray}
\mu_{x,y} & = & 2s_{x}-1 \\ \nu_{x,y} & = & 2r_{x}-1,
\end{eqnarray}
where $x=1,\ldots,N$.

%%%%%%%%%%%%%%%%%%%%%%%%%%%%%%%%%%%%%%%%%%%%%%%
\begin{figure}[h]
\includegraphics[width=8.5cm]{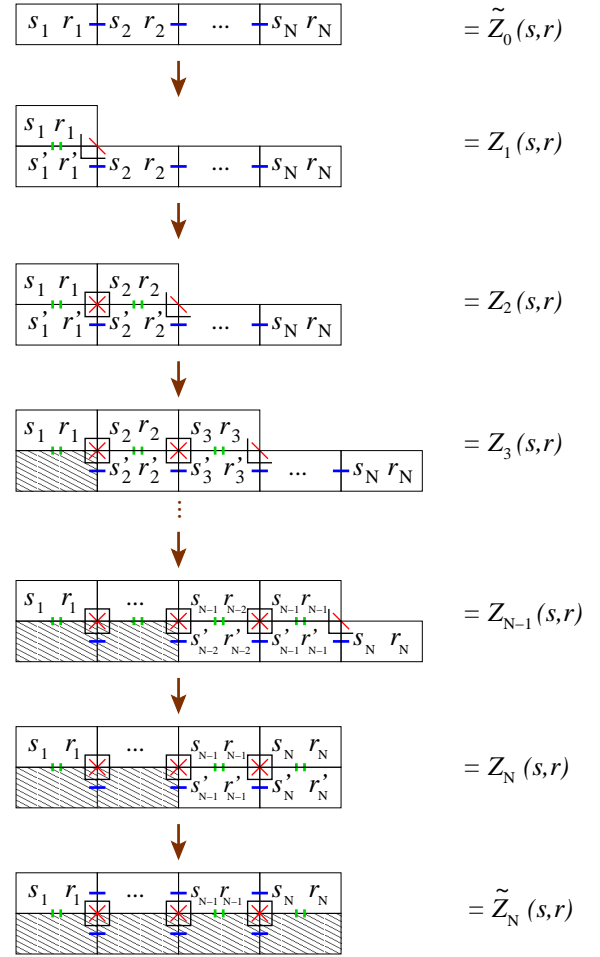}
\caption{(Color online) Thick horizontal lines represent two-body
  nearest-neighbor horizontal interactions. Short thick double
  vertical lines represent two-body nearest-neighbor vertical
  interactions. Diagonal thin lines represent two-body next-to-nearest
  neighbor interactions. L-shaped lines and squares represent
  three-body interactions.}
\label{fig:binder-method}
\end{figure}
%%%%%%%%%%%%%%%%%%%%%%%%%%%%%%%%%%%%%%%%%%%%%%%

Let $Z_{0}(\mathbf{s},\mathbf{r})$ denote the partition function term
containing the first (bottom) row of the lattice when only
nearest-neighbor interactions within that row are taken into
account. From this partial partition function we can build the full
partition function of the entire lattice (see Fig.
\ref{fig:binder-method}) with the following recursive protocol.
\begin{enumerate}
\item Find $Z_{0}(\mathbf{s},\mathbf{r})$ for the first (bottom) row.
\item Incorporate boundary fields (the dependence on boundary fields
  is left implicit hereafter):
\begin{equation}
\tilde{Z}_{0} (\mathbf{s}, \mathbf{r}) = Z_{0} (\mathbf{s},
\mathbf{r}) \kappa(\alpha_{b}, \beta_{b}, \mathbf{s}, \mathbf{r}).
\end{equation}
\item Evaluate $Z_{1} (\mathbf{s}, \mathbf{r}; s_{1}', r_{1}')$ at the
  second row, first site:
\begin{equation}
Z_{1} (\mathbf{s}, \mathbf{r}; s_{1}', r_{1}') = \eta( s_{1},
r_{1}; s_{2}, r_{2}, s_{1}', r_{1}') \tilde{Z}_{0} (\mathbf{s}',
\mathbf{r}'),
\end{equation}
where $\mathbf{s}' = \mathbf{s} + s_{1}' - s_{1}$ and $\mathbf{r}' =
\mathbf{r} + r_{1}' - r_{1}$.
\item Evaluate $Z_{2} (\mathbf{s}, \mathbf{r}; s_{2}', r_{2}')$ at the
  second row, second site:
\begin{eqnarray}
Z_{2} (\mathbf{s}, \mathbf{r}; s_{2}', r_{2}') & = & \eta
(s_{2},r_{2}; s_{3}, r_{3}, s_{2}', r_{2}') \nonumber \\ & & \times
\sum_{s_{1}',r_{1}'} \lambda( s_{1}, r_{1}; s_{2}, r_{2}; s_{1}',
r_{1}'; s_{2}', r_{2}') \nonumber \\ & & \times\ Z_{1} (\mathbf{s}',
\mathbf{r}'; s_{1}', r_{1}'),
\end{eqnarray}
where $\mathbf{s}' = \mathbf{s} + 2 (s_{2}'-s_{2})$ and $\mathbf{r}' =
\mathbf{r} + 2 (r_{2}'-r_{2})$.  Update $Z_{1} = Z_{2}$.
\item Evaluate $Z_{k} (\mathbf{s}, \mathbf{r}; s_{k}', r_{k};)$ at the
  second row, $k$-th site ($2 < k \leq N_{x}-1$):
\begin{eqnarray}
Z_{2} (\mathbf{s}, \mathbf{r}; s_{k}', r_{k}') & = & \eta( s_{k},
r_{k}; s_{k+1}, r_{k+1}, s_{k}',r_{k}') \sum_{s_{k-1}',r_{k-1}'}
\nonumber \\ & & \lambda( s_{k-1}, r_{k-1}; s_{k}, r_{k}; s_{k-1}',
r_{k-1}'; s_{k}', r_{k}') \nonumber \\ & & \times\ Z_{1} (\mathbf{s}',
\mathbf{r}';s_{k-1}', r_{k-1}'),
\end{eqnarray}
where $\mathbf{s}' = \mathbf{s} + 2^{k-1} (s_{k}'-s_{k})$ and
$\mathbf{r}' = \mathbf{r} + 2^{k-1} (r_{k}'-r_{k})$.  Update $Z_{1} =
Z_{2}$.
\item Evaluate $Z_{N_{x}} (\mathbf{s}, \mathbf{r})$ at the second row,
  $N_{x}$-th site:
\begin{eqnarray}
Z_{2}(\mathbf{s},\mathbf{r}) & = & \sum_{s_{N_{x}}',r_{N_{x}}'}
\gamma( s_{N_{x}}, r_{N_{x}}; s_{N_{x}}', r_{N_{x}}')
\sum_{s_{N_{x-1}}',r_{N_{x-1}}'} \nonumber \\ & & \lambda(
s_{N_{x}-1}, r_{N_{x}-1}; s_{N_{x}}, r_{N_{x}}; s_{N_{x}-1}',
r_{N_{x}-1}'; s_{N_{x}}', r_{N_{x}}') \nonumber \\ & & \times\ Z_{1}
(\mathbf{s}',\mathbf{r}'; s_{N_{x}-1}', r_{N_{x}-1}'),
\end{eqnarray}
where $\mathbf{s}' = \mathbf{s} + 2^{N_{x}-1}(s_{N_{x}}'-s_{N_{x}})$
and $\mathbf{r}' = \mathbf{r} + 2^{N_{x}-1}(r_{N_{x}}'-r_{N_{x}})$.
\item Evaluate $\tilde{Z}_{0}(\mathbf{s},\mathbf{r})$ to incorporate
  horizontal interactions in the second row:
\begin{equation}
\tilde{Z}_{0}(\mathbf{s},\mathbf{r}) = Z_{2}(\mathbf{s},\mathbf{r})
Z_{0}(\mathbf{s},\mathbf{r})
\end{equation}
\item Rename $\tilde{Z}_{0}(\mathbf{s},\mathbf{r})$ as
  $Z_{0}(\mathbf{s},\mathbf{r})$ and do another iteration (third row).
\item Repeat until, at the end of the $N_{y}-1$ iteration ($N_{y}$th
  row),
\begin{equation}
c(\alpha_{t},\alpha_{b},\beta_{t},\beta_{b}) =
\sum_{\mathbf{s},\mathbf{r}} \tilde{Z}_{0}(\mathbf{s},\mathbf{r})
\kappa(\alpha_{t},\beta_{t},\mathbf{s},\mathbf{r}).
\end{equation}
\end{enumerate}
The algorithm is straightforward to implement numerically once
expressions for the coefficients
%
%\begin{equation}
$\kappa(\alpha_{b},\beta_{b},\mathbf{s},\mathbf{r})$,
%\end{equation}
%
%\begin{equation}
$\eta(s_{x},r_{x};s_{x+1},r_{x+1};s_{x}',r_{x}')$,
%\end{equation}
%
%\begin{equation}
$\lambda(s_{x-1},r_{x-1};s_{x},r_{x};s_{x-1}',r_{-1}';s_{x}',r_{x}')$,
%\end{equation}
%
and 
%
%\begin{equation}
$\gamma(s_{N_{x}},r_{N_{x}};s_{N_{x}}',r_{N_{x}}')$
%\end{equation}
%
are provided. These coefficients incorporate boundary fields and
vertical nearest-neighbor interactions, as well as next-to-nearest
neighbor (diagonal) interactions and three-site interactions.

%%%%%%%%%%%%%%%%%%%%%%%%%%%%%%%%%%%%%%%%%%%%%%%%%%%%%%%%%%%%%%%%%%%%%%%%%%%%%%%

%%%%%%%%%%%%%%%%%%%%%%%%%%%%%%%%%%%%%%%%%%%%%%%%%%%%%%%%%%%%%%%%%%%%%%%%%%%%%%%


\begin{thebibliography}{29}
\bibitem{footnote} This assumption simply means that the same
  dispersion relation holds for all spatial dimensions. Although this
  is the most common situation, it is conceivable that in some
  solid-state environments it will not hold.
\end{thebibliography}
\end{document}